\documentclass[letterpaper,11pt]{article}
\pdfoutput=1
\usepackage{jheppub}
\usepackage{amsmath}
\usepackage{graphicx}% Include figure files
\usepackage{subcaption}
\usepackage{array}
\usepackage{slashed}
\usepackage{appendix}
\usepackage{multirow}
\usepackage{xcolor}
\usepackage{aas_macros} 
\newcommand{\skipnew}[1]{}

\title{\boldmath Putting the Brakes on Axion Strings: Friction and Its Impact on the QCD Axion Abundance}

\usepackage{tikz,xcolor,hyperref}
\definecolor{lime}{HTML}{A6CE39}
\DeclareRobustCommand{\orcidicon}{%
	\begin{tikzpicture}
	\draw[lime, fill=lime] (0,0) 
	circle [radius=0.16] 
	node[white] {{\fontfamily{qag}\selectfont \tiny ID}};	\draw[white, fill=white] (-0.0625,0.095) 
	circle [radius=0.007];	\end{tikzpicture}
	\hspace{-2mm}}
\foreach \x in {A,...,Z}{%
	\expandafter\xdef\csname orcid\x\endcsname{\noexpand\href{https://orcid.org/\csname orcidauthor\x\endcsname}{\noexpand\orcidicon}}
}
\author{Anson Hook,\orcidA}
\author{Rajrupa Mondal,\orcidB}
\author{and Shourya Mukherjee\orcidC}
\affiliation{Maryland Center for Fundamental Physics, Department of Physics,
University of Maryland, College Park, MD 20742-4111 USA}

\emailAdd{hook@umd.edu}
\emailAdd{rajondal@umd.edu}
\emailAdd{mshourya@umd.edu}

\abstract{
A compelling production mechanism for QCD axion dark matter is from the scaling dynamics of early universe axion strings. We show that in DFSZ-like models containing tree-level interactions between fermions and the axion, friction between the thermal bath and the axion string drastically changes the behavior of the axion string network for lower $f_a$ values. Friction delays the onset of scaling and increases the energy density of axions. Once the effects of friction are included, we argue that in addition to the standard value of $m_a \sim$ meV, $m_a \sim 0.1$ eV also reproduces the dark matter energy density.
}

\begin{document}
\maketitle
\flushbottom

\section{Introduction} \label{sec:intro}

Despite all of the great successes of the Standard Model (SM), it still does not explain all observed phenomena.  Two compelling reasons for the existence of Beyond the Standard Model (BSM) physics is the strong CP problem and dark matter.  The strong CP problem is the puzzle of why the neutron electric dipole moment (EDM) is so small, see e.g. Ref.~\cite{Hook:2018dlk} for a review.  Meanwhile dark matter is the puzzle that our universe is suffused with non-relativistic cold matter whose properties are not those of any of the observed particles.

The QCD axion provides a compelling solution to both of these problems~\cite{Peccei:1977hh,Weinberg:1977ma,Wilczek:1977pj}. The QCD axion is the pseudo-Nambu Goldstone boson of a Peccei-Quinn (PQ) symmetry, $U(1)_{\rm PQ}$, that is spontaneously broken at a scale $f_a$ and explicitly broken by its interaction with QCD.  It dynamically relaxes the neutron eDM to zero while it is weakly coupled enough to play the role of dark matter~\cite{Preskill:1982cy,Abbott:1982af,Dine:1982ah}. Additionally, the QCD axion has two very compelling production mechanisms, the pre- and post-inflationary Peccei-Quinn (PQ) symmetry breaking scenarios.  In the first scenario, inflationary fluctuations push the QCD axion to a random initial field value.  After inflation, this initial field value is automatically converted into particles via the misalignment mechanism~\cite{Preskill:1982cy,Dine:1982ah,Abbott:1982af}.

The post-inflationary PQ symmetry breaking scenario occurs when the universe is reheated to temperatures above the axion decay constant $f_a$, where the PQ symmetry is restored.  As the universe cools through the PQ phase transition, axion strings form via the Kibble mechanism~\cite{Kibble:1976sj,Zurek:1985qw} and persist until the QCD phase transition, when domain walls appear. For domain wall number $N_{\rm DW}=1$~\cite{Sikivie:1982qv,Vilenkin:1984ib,Hiramatsu:2010yn,Hiramatsu:2012gg}, the network annihilates. The energy radiated by this network of strings and domain walls goes into axions and is how this scenario produces axion dark matter. Because there is only one free parameter in this production mechanism, $f_a$, there is only a single value of $f_a$, or equivalently the axion mass, that produces the observed dark matter abundance.  This prediction is another one of the beautiful features of this production mechanism.

Due to the potential for a precise prediction of the mass of axion dark matter, there has been much numerical work to understand the production of axions from strings~\cite{Gorghetto:2018myk,Gorghetto:2020qws,Klaer:2017ond,Hindmarsh:2017qff,Buschmann:2019icd,Vaquero:2018tib,Fleury:2015aca,Kawasaki:2018bzv}. These simulations have found that the string networks approach a scaling regime with logarithmic violations. Because obtaining the QCD axion abundance requires extrapolations to a hierarchy of length scales well beyond current numerical reach, these numerical simulations have not yet arrived at a consensus for the value of the predicted dark matter axion mass.

One important piece of physics that is not captured by numerical simulations is the interaction between the string and the ambient plasma, which can manifest itself as a frictional effect.  The interaction between axion strings and gluons due to the defining coupling of the QCD axion, $\frac{g^2}{32 \pi^2} \frac{a}{f_a} G \tilde G$, has been calculated in Ref.~\cite{Agrawal:2020euj}.  It was found that due to the $\frac{g^2}{32 \pi^2}$ suppression in the coupling, that the effects of friction are subdominant at the QCD phase transition when most of the dark matter axions are produced.  It is thus justified to ignore friction in models, such as the KSVZ model, that only generate the topological coupling to the gauge bosons of the form $a G \tilde G$~\cite{Kim:1979if,Shifman:1979if}.

The situation is vastly different for models that also generate tree level couplings with fermions, such as the DFSZ model~\cite{Zhitnitsky:1980tq,Dine:1981rt} or any model which connects the axion with flavor~\cite{Davidson:1981zd,Wilczek:1982rv,Davidson:1983fy,Ema:2016ops,Calibbi:2016hwq,Arias-Aragon:2017eww,Bjorkeroth:2018dzu}. These models contain interactions of the form $\mathcal{O}(1) \frac{\partial_\mu a}{f_a} \overline \psi \gamma^\mu \gamma^5 \psi$, which can generate large frictional effects due to Aharonov-Bohm scattering~\cite{Aharonov:1959fk,Alford:1988sj}.  One can easily see that these unsuppressed interactions\footnote{The $1/f_a$ in the interaction is canceled because the axion changes by $2 \pi f_a$ around an axion string.} can change the dynamics of axion strings by comparing the curvature induced force per unit length on a Hubble sized string segment, $f_a^2 H $, against the effects of friction, $T^3$.  For any $f_a \lesssim 10^9$ GeV, friction will be important at temperatures of order the QCD scale.  

In this article, we demonstrate that the effect of friction in DFSZ like models is to drastically increase the production of low-$f_a$ axions such that axions with $f_a \sim 10^8$ GeV, $m_a \sim 0.1$ eV, can also reproduce the observed abundance of dark matter. This is a particularly interesting parameter region as axions with this mass are either on the verge of discovery~\cite{CAST:2008ixs,CAST:2011rjr,CAST:2013bqn,CAST:2015qbl} or in slight tension with existing astrophysical constraints~\cite{Ayala:2014pea,XENON:2020rca,Capozzi:2020cbu,Dolan:2022kul}. To demonstrate the viability of this data point most convincingly, one would need to incorporate frictional effects on strings at temperatures of order the QCD confinement scale.  Given that this simulation is yet to be done for frictionless strings, it is not surprising that the frictional case is also beyond current capabilities.  

Instead, we develop a semi-analytical framework utilizing the velocity-one-scale (VOS) model~\cite{Martins:1996jp,Martins:2000cs} rescaled in such a way that the frictionless VOS predictions agree with existing frictionless numerical simulations.   We translate the resulting string network history into axion production using two complementary estimators based on two separate methods used by existing numerical simulations.  In the first approach we use a loop-based log-uniform loop spectrum with friction-limited radiating sizes, while in the second approach we use a spectrum-based method that adapts the instantaneous emission prescription of large-volume simulations with a friction-induced infrared cutoff.  Finally, following Ref.~\cite{Gorghetto:2020qws}, we include the nonlinear QCD-era evolution between the onset of oscillations and the end of the relativistic phase by tracking the infrared gradient energy until it becomes comparable to the axion potential energy. 

The paper is organized as follows. In Sec.~\ref{Sec: review}, we review the velocity-dependent one-scale model predictions for the behavior of a string network experiencing friction. In Sec.~\ref{sec:SemiAnalytic}, we describe the framework of our semi-analytical approach to modeling string networks with friction. In Sec.~\ref{subsec:method1}, and Sec.~\ref{subsec:Method2}, we discuss two numerically inspired approaches to axion emission and their implications for axion dark matter.  
Sec.~\ref{Sec:conclusion} concludes the paper. In App.~\ref{App:DW}, we discuss the effects of domain walls.  In App.~\ref{App:fric}, we calculate the friction on axion strings coming from axion couplings to fermions.  Finally in App.~\ref{App:VOS}, we review the velocity one-scale model.

\section{Dynamics of a Frictionfull String-Loop Network} \label{Sec: review}

In this section, we discuss the behavior of a string-loop network in the presence of friction.  Before addressing the impact of friction, let us first summarize the results for the frictionless case.  As the temperature of the universe falls, eventually the global $\rm{U}(1)_{\rm{PQ}}$ is spontaneously broken at a temperature $T\sim f_a$ producing strings.  While the initial distribution of strings depends sensitively on the dynamics of the PQ phase transition, the string network quickly approaches an attractor solution called the scaling regime~\cite{Vilenkin:2000jqa,Kibble:1976sj}.

The scaling regime is an approximately scale-invariant attractor solution characterized by a roughly constant number of long strings per Hubble volume. During the scaling regime, string energy is constantly being redistributed between closed string loops and long strings by intercommutation and fragmentation. Energy is lost from the network by radiation into axions. Finally, expansion of the universe is constantly stretching out the strings (increasing their total energy) and diluting them away (decreasing their total energy density). The interplay among all of these effects gives rise to the scaling regime.

Typically, the scaling regime is parameterized using a dimensionless quantity $\xi_{\rm tot}(t)$ equal to the total string length per Hubble volume~\cite{Gorghetto:2018myk},
\begin{align}
    \xi_{\rm tot}(t) \equiv \lim_{V\to\infty} \frac{\ell_{\rm tot}(V)\, t^{2}}{V}\,,
    \label{eq:xi}
\end{align}
where $\ell_{\rm tot}(V)$ is the total string length contained in a volume $V$.  Meanwhile, the total energy density in strings with tension $\mu(t)$ is
\begin{align}
    \rho_{s, \rm tot}(t) = \xi_{\rm tot}(t)\,\frac{\mu(t)}{t^{2}}\,.
    \label{eq:rhos_def}
\end{align}
For global strings, the tension is enhanced by the long-range Goldstone field and is well approximated by~\cite{Gorghetto:2018myk,Klaer:2017qhr,PhysRevLett.48.1867}
\begin{equation}
  \mu(t) \simeq \pi f_a^2\, \ln N(t)\,,
  \qquad
  \ln N(t) \equiv \ln\left(\frac{m_r}{\sqrt{\xi_{\rm tot}(t)} H(t)}\right),
  \label{eq:tension_def}
\end{equation}
where $H(t) = 1/2t$ is the Hubble rate during radiation domination, and $m_r \sim f_a$ is the radial (core) mass scale of the PQ field set by the symmetry-breaking scale. The logarithm varies slowly with time and typically reaches values $\ln N = \mathcal{O}(50-70)$ at the relevant late-time cosmological epochs. The defining feature of the scaling regime for global strings is that
\begin{equation}
  \xi_{\rm tot}(t) = c_1 \ln N(t) + c_0 \, ,
\end{equation}
where $c_1$ and $c_0$ are found by fitting to numerical simulations~\cite{Gorghetto:2018myk,Gorghetto:2020qws}.

Loops are produced when long strings intersect and reconnect, chopping off segments whose initial sizes are a fraction of the correlation length $L$. Additional loops arise from the fragmentation of parent loops as kinks and cusps develop~\cite{Vilenkin:2000jqa}. In the large log regime relevant for global strings, axion radiation is the dominant channel for energy loss, with emission sourced by oscillating loops and by small-scale structure on long strings. The mean energy of quanta emitted by a loop of length $\ell$ scales as $\omega\sim 1/\ell$~\cite{Vilenkin:2000jqa}. 

Let us now discuss how adding friction changes the dynamics of the string network.  The discussion will be an overview with the details relegated to App.~\ref{App:fric} and App.~\ref{App:VOS}. A frictionfull string network is characterized by three qualitatively different regimes: (i) an initial stretching phase, during which strings are at rest and are being stretched by the expansion of the universe; (ii) a transient regime (the Kibble regime), in which both friction and the straightening of strings due to tension are important; and (iii) the scaling regime, where friction is unimportant.

The effects of a plasma are encoded in the drag force per unit length
\begin{equation}
  \frac{F_{\rm drag}}{\textit{l}} \;=\;
  -\frac{\mu(t)}{\ell_f(t)} \, \gamma(t) \,v(t)
\end{equation}
where $v(t)$ is the rms velocity of the long-string, $\gamma(t)$ is the Lorentz factor and $\ell_f$ is an effective friction length. The quantity  $\ell_f(t)$ is  useful, as any string  with radius of curvature $R$  smaller than $\ell_f$ will be essentially friction-free, while any string with $R \gtrsim \ell_f$ is dominated by friction. The dominant source of friction in our analysis is axial Aharonov-Bohm (AB) scattering of relativistic SM fermions off the string, induced by the derivative axion-fermion coupling. The AB contribution scales as
\begin{equation}
  \ell_{f,{\rm AB}}(T) \propto \frac{\pi f_a^2 \ \ln N}{T^3}\,
  \bigg[\sum_f g_f(T)\,\sin^2\Big(\frac{\pi C_f}{2}\Big)\bigg]^{-1} \,,
  \label{eq:lf_scaling_text}
\end{equation}
where $C_f$ are axial charges defined in Eq.~\eqref{Eq: axial coupling} and $g_f(T)$ counts the relativistic fermionic degrees of freedom~\cite{Alford:1988sj, Vilenkin:1984ib}. From Eq.~\eqref{eq:lf_scaling_text} one sees that smaller $f_a$ leads to stronger damping, since $\ell_{f,{\rm AB}}\propto f_a^2$, and that friction weakens rapidly (compared to Hubble) as the universe cools, with $\ell_{f,{\rm AB}}\propto T^{-3}$. Additional microphysical processes, such as scattering via core interactions with scalars~\cite{Vilenkin:2000jqa}, AB scattering of gauge bosons~\cite{Agrawal:2020euj}, and possible electromagnetic drag for superconducting strings~\cite{Witten:1984eb,Ostriker:1986xc}, are parametrically subleading, though their effects can also be included in the total friction length $\ell_f(T)$. We also neglect the radiative backreaction term sometimes added to the long-string VOS evolution, $\propto v^6$, since it only induces a subleading correction to the macroscopic network dynamics in the regime of interest~\cite{Martins:2000cs,Battye:1993jv}. In what follows, the AB contribution in Eq.~\eqref{eq:lf_scaling_text} is the dominant source of damping, and we refer to App.~\ref{App:fric} for further details.

To describe the behavior of a frictionfull string network, we utilize the velocity-dependent one-scale (VOS) model~\cite{Martins:1995tg,Martins:2000cs}. The idea behind the VOS model is to assume that the string network is characterized entirely by a single length scale and a single velocity scale. To derive the VOS equations, one first approximates an axion string as an object described by the Nambu-Goto effective action, despite its extended logarithmic structure.  One can then show that the correlation length $L(t)$ and the rms velocity $v(t)$ of the long strings obey
\begin{align}
  2\,\frac{dL}{dt} &= 2HL(1+v^2) + c\,v + \frac{L v^2}{\ell_f}\,,
  \label{eq:VOS_L_eq}\\[4pt]
  \frac{dv}{dt} &= (1-v^2)\left[\frac{k_v}{L} - v\left(2H + \frac{1}{\ell_f}\right)\right],
  \label{eq:VOS_v_eq}
\end{align}
where $k_v \equiv k(v)$ encodes the average curvature of long strings, and $c$ is the loop chopping efficiency that parametrizes energy loss into loops. In the three regimes, these equations admit simple asymptotic solutions. Denoting by $L_c$ the initial correlation length at the string network formation time $t_c$, one finds schematically
\begin{equation}
  L(t) \ \sim\ 
  \begin{cases}
    L_c \left( \dfrac{t}{t_c} \right)^{1/2} & \text{stretching regime} , \\[8pt]
    \displaystyle \sqrt{\dfrac{2(k_v+c)}{3\theta}}\ 
    \dfrac{t^{5/4}}{t_c^{1/4}}\,\sqrt{\ln N} & \text{Kibble regime}, \\[10pt]
    \sqrt{k_v(k_v+c)}\  t  & \text{scaling regime}\,,
  \end{cases}
  \label{eq:L}
\end{equation}
where $\theta$ (defined in Eq.~\eqref{eq:theta_def}) parametrizes the effective friction strength. A detailed derivation of these solutions is given in App.~\ref{App:VOS}.

The main effects of friction can be understood qualitatively as follows. In the early Kibble regime, where $\ell_f \ll L$, the drag term dominates the velocity equation and drives $v\ll1$, so long strings are effectively frozen into the plasma and stretched by the Hubble flow. This suppresses the growth of small-scale structure and loop production, and friction acts as an effective stiffness that resists bending and chopping. As the Universe cools, $\ell_f \propto T^{-3}$ grows and a given network element that was overdamped at early times becomes free to move and radiate once $\ell_f$ exceeds its physical size. Consequently, a significant fraction of axion emission is shifted closer to the onset of oscillations, defined by $H(t_{\rm osc})\simeq m_a(t_{\rm osc})$.

It is useful to compare this with the effect of the string tension. The tension $\mu$ tends to straighten the network dynamically by accelerating strings away from curved configurations, while friction suppresses the effects of curvature by damping their motion. Thus both effects have similar results (in different $f_a$ regions) because they affect the axion production dynamics heuristically in a similar manner. In the former, larger $\mu$ (larger $f_a$) promotes an earlier approach to scaling, whereas stronger friction at small $f_a$ delays efficient emission and thus increases the surviving axion abundance.

\section{A Semi-Analytical Framework}\label{sec:SemiAnalytic}

A faithful treatment of friction in post-inflation axion-string networks would require large, real-time simulations that evolve the string cores, the long-range Goldstone field, and the thermal plasma together through the QCD epoch and over many Hubble times. Even without friction, current simulations -- either large-volume lattice runs with controlled extrapolations in the large logarithm $\ln(m_r/H)$~\cite{Kawasaki:2018bzv,Gorghetto:2018myk,Gorghetto:2020qws}, or adaptive-mesh-refinement (AMR) studies~\cite{Buschmann:2019icd,Buschmann:2021sdq,Benabou:2024msj} -- are technically demanding, computationally expensive, and cannot resolve all the relevant length scales. The main challenges are the huge dynamic range between the core scale $m_r^{-1}$ and the Hubble radius $H^{-1}$, the need to resolve small-scale structure on strings while covering horizon-sized volumes, and the rapid growth of the axion mass near $T_{\rm QCD}$~\cite{Borsanyi:2016ksw}, which further complicates the problem.

Implementing friction significantly elevates this challenge. First, one must couple the axion sector to a multi-component relativistic plasma and account for the various scattering processes of SM particles off the strings. Second, a consistent cosmological treatment must evolve temperature-dependent transport coefficients, the axion mass $m_a(T)$ near the QCD crossover, and the changing effective degrees of freedom $g_f(T)$. Implementing all these ingredients at the spatial and temporal resolutions needed to control the large log, $\ln N$, is beyond current computational capabilities and would require a new class of plasma-string hybrid simulations.

Given these limitations, we adopt a semi-analytical strategy. We evolve the axion-string network with the VOS model, but we calibrate it to existing frictionless simulations, and we add friction through a microphysical damping length, $\ell_{f}(T)$, via Eq.~\eqref{eq:VOS_L_eq}. It is important to stress that in the VOS framework $\xi(t)$ is defined via the energy density of \emph{long strings} only, $\rho_s = \xi(t)\,\mu/t^2$, whereas numerical simulations often report a parameter $\xi_{\rm tot}$ extracted from the \emph{total} string length (long strings plus loops), via the total string energy density $\rho_{s,{\rm tot}}=\xi_{\rm tot}(t)\,\mu/t^2$. Frictionless scaling simulations indicate that the energy is split approximately $1 \, {:} \, 4$ between loops and long strings~\cite{Gorghetto:2018myk,Klaer:2017qhr}, so that $\rho_s \simeq \frac45 \,\rho_{s,\rm tot}$ and we may identify $\xi(t) \simeq \frac45 \,\xi_{\rm tot}(t)$ in the scaling regime. With this convention, the framework returns the long-string density $\xi(t)$, the correlation length $L(t)\simeq t/\sqrt{\xi(t)}$, and the rms velocity $v(t)$ across the Kibble (friction-dominated) epoch and into the transition toward scaling.

To set the normalization of the long-string energy density, we rescale the entire VOS solution for $\xi(t)$ by a constant factor chosen to match large-volume lattice simulations in the relativistic scaling regime,
\begin{equation}
  \xi(t) =
  \frac{\frac45 \ \xi_{\rm num}^{\rm scaling}(t)}{\xi_{\rm VOS}^{\rm scaling}} \ 
  \xi_{\rm VOS}(t),
  \label{eq:strategy_rescale}
\end{equation}
where 
\begin{equation}
  \xi_{\rm num}^{\rm scaling}(t) = \frac45\,\xi_{\rm num,tot}^{\rm scaling}(t) \simeq \frac45\,\bigl(c_1 \ln N(t) + c_0\bigr)
\end{equation}
denotes the long-string scaling value inferred from frictionless lattice simulations of global strings. These simulations find $c_1 \approx 0.21-0.25$, with values $0.24(2)$~\cite{Gorghetto:2020qws}, $0.22(2)$~\cite{Gorghetto:2018myk}, $0.21(2)$~\cite{Benabou:2024msj}, $0.254(2)$~\cite{Buschmann:2021sdq}, and $0.23(6)$~\cite{Saikawa:2024bta}. Most results are insensitive to the subleading constant term $c_0 = \mathcal{O}(1)$ in the $\ln N$ expansion.

$\xi_{\rm VOS}^{\rm scaling}$ is the corresponding scaling value predicted by the frictionless VOS model. $\xi_{\rm VOS}(t)$ denotes the full time-dependent VOS solution including friction across all regimes as derived in App.~\ref{App:VOS}. In the limit where friction becomes negligible, the VOS solution reduces to the standard frictionless system and the rescaled $\xi(t)$ reproduces, by construction, the scaling measured in lattice simulations.

We convert the network evolution into axion number production using two complementary estimators that emphasize different aspects of the same physics and help bound modeling systematics. In both cases, the physical axion number density at the onset of oscillations, $t_{\rm{osc}}$, obtained by integrating the instantaneous production rate, with each contribution redshifted to $t_{\rm{osc}}$,
\begin{equation}
  n(t_{\rm{osc}})
  =
  \int_{t_K}^{t_{\rm{osc}}} d\tilde t\
  \left(\frac{dn}{d\tilde t}\right)_{\rm{prod}}
  \left[\frac{a(\tilde t)}{a(t_{\rm{osc}})}\right]^{3}
  \label{eq:redshift_to_t0}
\end{equation}
where $(dn/d\tilde t)_{\rm{prod}}$ is the instantaneous axion production rate per physical volume at time $\tilde t$, and $t_K$ marks the onset of the Kibble regime defined as Eq.~\eqref{eq:tK}. The factor $[a(\tilde t)/a(t_{\rm osc})]^{3}$ accounts for dilution from cosmic expansion. The two estimators differ only in how they model $(dn/d\tilde t)_{\text{prod}}$.\\

The first method, described in detail in Sec.~\ref{subsec:method1}, assumes an approximately log-uniform instantaneous loop spectrum. Since friction prevents large loops from oscillating efficiently, we take only loops with $\ell \lesssim \ell_{\rm crit}(t)$ to radiate, where the critical size is set by the smaller of the friction length and the maximal loop size at formation,
\begin{equation}
  \ell_{\rm crit}(t) \sim \min \Big[\ \ell_f(t) , \ L(t) \ \Big]
  \label{eq:lcrit_def}
\end{equation}
and we then use the nearly length-independent loop power, $P_a(\ell) \propto f_a^2$, to obtain a closed expression for the instantaneous axion number injection. Friction enters only through $\ell_{\rm crit}$, which switches from $\ell_f$ (early Kibble) to $L$ (late Kibble/near-scaling). 

The second method, discussed in Sec.~\ref{subsec:Method2}, is spectrum-based. Following the prescription of Ref.~\cite{Gorghetto:2018myk}, we integrate the normalized emission spectrum $F \big(k/H,\,m_r/H\big)$ over the kinematic range allowed at time $t$, imposing a frictional infrared cutoff
\begin{equation}
  k_{\min}(t) \sim \max\Big[ \ \ell_f^{-1}(t),\,L^{-1}(t)\ \Big]\,.
  \label{eq:kmin_def}
\end{equation}
In this formulation, friction only enters through the time-dependent lower limit of integration, accounting for the fact that modes with wavelength larger than $\ell_f(t)$ do not oscillate and therefore do not contribute. The ultraviolet cutoff is set by the core scale $m_r$. A full definition of the spectral shape and its normalization is deferred to Sec.~\ref{subsec:Method2}.

Finally, we compute the axion number density at the onset of oscillations, $n(t_{\rm osc})$ with $H(t_{\rm osc})\simeq m_a(t_{\rm osc})$. We track its subsequent evolution through the QCD-induced modification of the axion potential and the associated non-linear dynamics, as discussed in Sec.~\ref{sec:dilution}.

\section{Loop-based Estimator with Log-flat Loop Distribution} \label{subsec:method1}

In this section, we develop the loop-based estimator introduced above. Motivated by large-scale simulations of axion-string networks, we model the instantaneous loop-size distribution as log-uniform up to the correlation length $L(t)$,
\begin{equation}
  \frac{d n_\ell(t)}{d\ell} = \frac{\mathcal{N}(t)}{\ell} \, ,
  \label{eq:logUniform}
\end{equation}
where $\mathcal{N}(t)$ is the normalization coefficient. We assume that at any given time, a fixed fraction $f_{\rm loop}$ of the long-string energy density resides in loops~\cite{Gorghetto:2018myk,Klaer:2017qhr},
\begin{equation}
  \rho_{\ell}(t) \equiv f_{\rm loop}\,\rho_s(t)\,.
\end{equation}
The same quantity can be expressed directly in terms of the loop distribution. For loops of length $\ell$, the energy per loop is $\mu(t)\,\ell$, so for an ensemble with number density $d n_\ell(t)/d\ell$ one has
\begin{align}
  \rho_{\ell}(t) 
  &= \int_0^{L(t)} d\ell\,\frac{d n_\ell(t)}{d\ell}\,\mu(t)\,\ell
   = \mu(t)\,\mathcal{N}(t)\,L(t)\,.
\end{align}
Using the long-string energy density $\rho_s(t) = \xi(t)\,\mu(t)/t^2$ and $L(t) = t/\sqrt{\xi(t)}$ then fixes the normalization as 
\begin{equation}
  \mathcal{N}(t) = f_{\rm loop}\,\frac{\xi^{3/2}(t)}{t^3}\,.
  \label{eq:N_of_t}
\end{equation}
Throughout, $\xi(t)$ denotes the rescaled VOS solution defined in Eq.~\eqref{eq:strategy_rescale}. As discussed above, frictionless simulations indicate a roughly $1\, {:}\,4$ split of the total string energy between loops and long strings, corresponding to $f_{\rm loop} = 0.25$, consistent with the range $f_{\rm loop}\sim 0.1$--$0.3$ reported in Refs.~\cite{Gorghetto:2018myk,Klaer:2017qhr}. Operationally, $f_{\rm loop}$ parametrizes the efficiency with which long strings self-intersect and chop into loops, which more generally should be regarded as a nuisance parameter encoding the loop-production efficiency, which can be varied within the simulation-motivated range.

For global strings at large logarithm, the axion power emitted by a loop is approximately independent of its length,
\begin{equation}
  \frac{dE_a}{dt}\Big|_\ell \simeq \Gamma_a f_a^2\,,
\end{equation}
where $\Gamma_a=\mathcal{O}(10^2)$ encodes spectral details~\cite{Gorghetto:2018myk,Fleury:2015aca,Chang:2021afa,Buschmann:2021sdq}. The typical momentum of the emitted quanta scales as $k\sim 1/\ell$, so the axion number emission rate of a loop of length $\ell$ is
\begin{equation}
  \left.\frac{dN_a}{dt}\right|_\ell \simeq \frac{(dE_a/dt)_\ell}{k} \simeq \Gamma_a f_a^2\,\ell\,.
  \label{eq:numberRate}
\end{equation}
Friction limits which loops can radiate efficiently, and we encode this by restricting the instantaneous emission to loops with length below a time-dependent critical size $\ell_{\rm crit}(t)$. In the late scaling regime, frictionless simulations impose an infrared cutoff on the largest radiating modes, $k_{\rm IR} = x_{\rm IR} H$ with $x_{\rm IR} \sim 30\text{--}100$~\cite{Buschmann:2021sdq,Benabou:2024msj}, which we map onto a cutoff on the physical loop size.

Consider a circular loop of radius $r_{\rm crit}$ and length $\ell = 2\pi r_{\rm crit}$. The longest oscillation that still radiates efficiently can be thought of as a disturbance that travels across the diameter and back in one period, corresponding to a wavelength of order $4 r_{\rm crit}$ and hence a minimal wavenumber $k_{\min} \sim 1/4 r_{\rm crit}$. Identifying this with the simulation $k_{\min} = k_{\rm IR}$ gives
\begin{equation}
  \ell_{\rm crit}^{\, \rm sc}(t) \simeq
  \frac{\pi}{x_{\rm IR}}\,t  \simeq
  \frac{\pi}{x_{\rm IR}}\sqrt{\xi(t)}\,L(t)\,.
\end{equation}
In the friction-dominated regime, loops larger than the friction length $\ell_f(t)$ are overdamped and do not radiate efficiently. As a fiducial estimate we take the critical size to be proportional to $\ell_f(t)$ and parametrize the associated $\mathcal{O}(1)$ uncertainty by
\begin{equation}
    \ell_{\rm crit}^{\, \rm K}(t) \equiv \alpha\,\ell_f(t)\,,
\end{equation}
with $\alpha \simeq 1 \text{--} 10$. In practice, we interpolate the restriction on the instantaneous emission to loops as 
\begin{equation}
    \ell \leq \ell_{\rm crit}(t) = \rm{min} \Big[ \, \ell_{\rm crit}^{\, \rm K}(t) \ , \ \ell_{\rm crit}^{\, \rm sc}(t)  \, \Big] \, .
\end{equation}
The instantaneous axion number injection rate per physical volume at time $\tilde t$ is then
\begin{align}
  \left.\frac{dn(\tilde t)}{d\tilde t}\right|_{\text{prod}}
  &= \int_{0}^{\ell_{\rm crit}(\tilde t)} d\ell\,
     \frac{d n_\ell(\tilde t)}{d\ell}\,
     \left.\frac{dN_a}{d\tilde t}\right|_\ell \nonumber\\
  &= \Gamma_a f_a^2\,\mathcal{N}(\tilde t)\,\ell_{\rm crit}(\tilde t)\,,
  \label{eq:injection}
\end{align}
with $\mathcal{N}(\tilde t)$ given by Eq.~\eqref{eq:N_of_t}. Friction enters through $\ell_{\rm crit}(\tilde t)$, which interpolates from $\ell_f(\tilde t)$ in the strongly damped Kibble regime to $L(\tilde t)$ once the network approaches scaling.

Substituting Eq.~\eqref{eq:injection} into the general redshifting formula Eq.~\eqref{eq:redshift_to_t0}, and using $a(t)\propto t^{1/2}$ in radiation domination, we obtain
\begin{equation}
  n(t_{\rm osc})
  = f_{\rm loop}\,\Gamma_a f_a^2\,
    \frac{1}{t_{\rm osc}^{3/2}}
    \int_{t_K}^{t_{\rm osc}} d\tilde t\,
    \frac{\xi^{3/2}(\tilde t)}{\tilde t^{3/2}}\,
    \ell_{\rm crit}(\tilde t)\,,
  \label{eq:n_radiation}
\end{equation}
which makes explicit the competition between the evolving network morphology (through $\xi$ and $L$) and dissipative microphysics (through $\ell_f$).

For presentation, it is convenient to compare directly to the dark-matter number density required to produce the observed energy density in dark matter today. We therefore define
\begin{equation}
  \Omega_r
  \equiv
  \frac{n(t_{\rm osc})}{n_{\rm DM}^{\rm req}(t_{\rm osc})}\,,
  \qquad
  n_{\rm DM}^{\rm req}(t)\equiv
  \frac{\rho_{c,0}\,\Omega_c}{m_a^0}
  \left(\frac{a_0}{a(t)}\right)^3,
  \label{eq:OmegaR_DM_here}
\end{equation}
where $\rho_{c,0}$ is the present-day critical density, $\Omega_c$ is the present-day cold dark matter fraction, and $m_a^0$ is the zero-temperature axion mass~\cite{GrillidiCortona:2015jxo}. We take
\begin{equation}
    m_a^0 \;\simeq\; 5.7\times 10^{-6} \ \mathrm{eV}
  \left(\frac{10^{12}\,\mathrm{GeV}}{f_a}\right),
\end{equation}
so that $\Omega_r$ directly measures the axion abundance produced by the this method relative to the number density required for cold dark matter. \\

\begin{figure}[t!]
  \centering
  \includegraphics[width=0.90\linewidth]{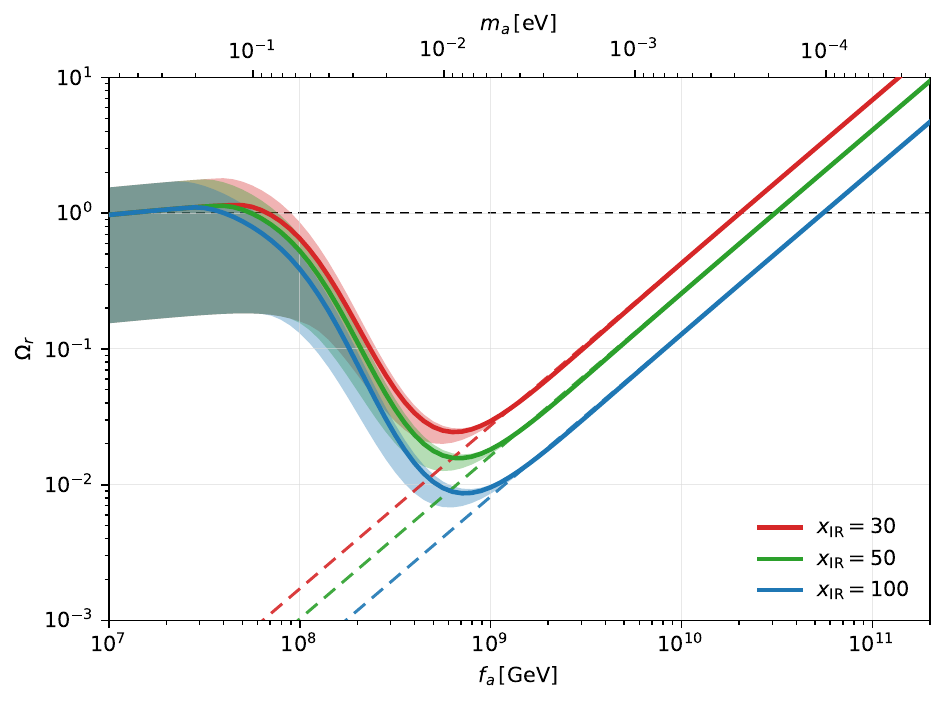}
  \caption{ Impact of friction on the ratio of axion abundance as computed by the loop-based estimator Eq.~\eqref{eq:n_radiation}. The shaded band shows the variation in the nuisance parameter $\alpha$ that sets the friction-limited critical loop size $\ell^K_{\rm crit} = \alpha\,\ell_f$, with $\alpha \in [1,10]$, while the solid curve corresponds to $\alpha = 2\pi$. We fix the loop emission coefficient to $\Gamma_a = 50$ and the loop fraction to $f_{\rm loop} = 0.25$. Different colors correspond to the infrared spectral cutoff $x_{\rm IR} = 30,50,100$. In the friction-limited regime at small $f_a$, the predicted axion abundance is essentially independent of $x_{\rm IR}$, whereas in the relativistic scaling regime at large $f_a$ the choice of $x_{\rm IR}$ sets the location of the $\Omega_r \simeq 1$ crossing, which lies in the range $f_a \sim 3\times 10^{10}\text{--}10^{11}\,\mathrm{GeV}$ in agreement with Refs.~\cite{Buschmann:2021sdq,Benabou:2024msj}. The dashed lines show the corresponding predictions when friction is neglected; including friction enhances the axion abundance by a factor of $\sim 10^1\text{--}10^4$, depending on the value of $f_a$ within the friction-dominated regime.}
  \label{fig:LogUniform}
\end{figure}

In Fig.~\ref{fig:LogUniform}, we show the result of the loop-based estimator in Eq.~\eqref{eq:n_radiation}, plotted as $\Omega_r$ versus $f_a$. The solid curve corresponds to a benchmark value of $\alpha = 2\pi$, while the shaded band is obtained by varying $\alpha \in [1-10]$ over the range compatible with current simulations. In this plot, there are three distinct regimes: \\

\textbf{Scaling regime emission} --  For large decay constants, $f_a \gtrsim 10^9\,\mathrm{GeV}$, the string network has already reached its relativistic scaling attractor well before $t_{\rm osc}$, with $\xi(t) = \xi_{\rm num}^{\rm scaling}(t)$ approaching a nearly constant value. In this regime, friction is negligible and  $\ell_{\rm crit}(t)\simeq L(t)$ throughout the epoch that dominates axion production. The axion emission is therefore controlled by the scaling configuration of the network, with long-string energy density $\rho_s(t) \propto f_a^2$. Since the required dark matter number density $n_{\rm DM}^{\rm req}(t)$ also increases with $f_a$, the net result in this high-$f_a$ regime is that, as one moves from large to smaller $f_a$, the predicted relic abundance $\Omega_r$ falls approximately as a power law.\\

\textbf{Correlation length-limited emission} -- For intermediate decay constants, $6 \times 10^7\,\mathrm{GeV} \lesssim f_a \lesssim 10^9\,\mathrm{GeV}$, the network spends a sizable fraction of the relevant epoch in a morphology-limited Kibble regime and only approaches the relativistic scaling solution around $t_{\rm osc}$. In this window $\xi(t)$ is larger than its scaling value and is still changing appreciably at $t_{\rm osc}$.  It is roughly inversely proportional to $f_a$. Meanwhile, the emission remains morphology-limited with $\ell_{\rm crit}(t)\simeq L(t)$.   The combined effect is for $\Omega_r$ to increase as $f_a$ decreases, rather than the power-law fall characteristic of the high-$f_a$ scaling regime. This effect can be seen by the departure of the solid line from the scaling prediction (dashed line).\\

\textbf{Friction-limited emission} -- For small decay constants, $f_a \lesssim 6 \times 10^7\,\mathrm{GeV}$, the QCD phase transition occurs while the network is still deep in the Kibble regime, so the strings remain strongly friction-damped around $t_{\rm osc}$. In this regime one has $\ell_{\rm crit}(t)\simeq \ell_f(t) < L(t)$, so axion emission is friction-limited rather than morphology-limited, and fewer total axions are emitted. Deep in the Kibble regime, $\xi(t) \propto f_a^{-2}$ while $\ell_f(t) \propto f_a^{5/2}$. Thus, as one moves from larger to smaller $f_a$ within this window, the number of long strings per Hubble volume increases (through $\xi$) but the friction length shrinks, reducing the range of radiating loop sizes. The smaller $\ell_f$ raises the effective IR momentum cutoff, which suppresses the emitted number density. Although the increase in $\xi$ partially compensates, the reduction in $\ell_f$ dominates, so $\Omega_r$ decreases as $f_a$ is lowered, producing the gentle downward trend.\\

From Eq.~\eqref{eq:n_radiation} one can identify the parameters that control the overall normalization and bandwidth of the band in Fig.~\ref{fig:LogUniform}. The abundance depends on the loop fraction $f_{\rm loop}$, the spectral factor $\Gamma_a$, the friction length $\ell_f(t)$, proportionality factor $\alpha$, and the VOS parameters $k_v$ and $c$ that determine the evolution of $\xi(t)$ and $L(t)$. Consequently, values $\Omega_r \gtrsim 1$ should be interpreted as indicating that the string network can readily produce enough axion dark matter, rather than as a precise prediction of the relic abundance.

\section{IR-dominated Spectral Estimator} \label{subsec:Method2}

In this method, we compute the axion number density at the onset of oscillations using a spectrum-based estimator that follows the prescription of Ref.~\cite{Gorghetto:2018myk,Gorghetto:2020qws} and adapt it to our string network system.

We describe the axion emission in terms of the instantaneous power injected into the axion energy density $\rho_a(t)$ per unit logarithmic interval of physical momentum $k$,
\begin{align}
    \frac{d\dot{\rho}_a}{d\ln k}(t)
    = \Gamma(t)\,\rho_{s,\rm tot}(t)\,
      F\left(x \equiv \frac{k}{H(t)},\ y \equiv \frac{m_r}{H(t)}\right),
    \label{eq:diff_energy_density}
\end{align}
where $\rho_{s,\rm tot}(t)$ is the total string energy density defined in Eq.~\eqref{eq:rhos_def}. The dimensionless function $F(x,y)$ describes the spectrum of axions emitted by the network at fixed $m_r/H$, while the prefactor introduced in Ref.~\cite{Gorghetto:2018myk},
\begin{equation}
    \Gamma(t) \simeq 2H \,-\, \frac{\dot{\xi}_{\rm tot}}{\xi_{\rm tot}}
\end{equation}
is an effective conversion rate per Hubble time that encodes how fast long-string energy is depleted into radiation.

In the stretching regime, the network is only conformally stretched, energy losses are negligible, so that $\Gamma_{\rm stretch} \simeq 0$. In the friction-dominated Kibble regime the long strings are strongly damped and their kinetic energy is continuously drained by interactions with the plasma. Using Eq.~\eqref{eq:L_Kibble} in the relation above one finds $\Gamma_{\rm K} \simeq 3H$. In the relativistic scaling regime, $\xi(t)\simeq\xi_{\rm num}^{\rm scaling}(t)$ is approximately constant (up to logarithmic corrections), so that $\Gamma_{\rm sc}(t) \simeq 2H\,$, consistent with the approximate scaling behavior observed in simulations.\\

For the spectral shape, we adopt the one-parameter family of Ref.~\cite{Gorghetto:2018myk}, parametrized by a spectral index $q$,
\begin{align}
    F(x,y) = \frac{1}{x_{\min}(t)}
    \left(\frac{x_{\min}(t)}{x}\right)^{q}\,
    \frac{q-1}{1-\left(\dfrac{x_{\min}(t)}{y}\right)^{q-1}},
    \qquad
    \int_{x_{\min}(t)}^{y(t)} d x\,F(x,y) = 1,
\label{eq:Fxy}
\end{align}
where the normalization condition is imposed over the kinematic range $\left[\, x_{\rm min}(t), y(t)\, \right]$. The lower limit
\begin{equation}
    x_{\min}(t) \;\equiv\; \frac{k_{\min}(t)}{H(t)}
\end{equation}
is the time-dependent infrared cutoff in units of the Hubble scale, with $k_{\min}(t)$ defined in Eq.~\eqref{eq:kmin_def}. For $q>1$, the spectrum is IR-dominated; the $q\to 1$ limit reproduces a log-flat $F\propto 1/x$ distribution, and $q<1$ gives a UV-dominated spectrum. The upper limit $y\equiv m_r/H\gg 1$ imposes the core cutoff, since modes with $k\gtrsim m_r$ are exponentially suppressed by the finite thickness of the string.

Friction and network morphology enter the spectrum through the infrared cutoff, which is set by the largest physical length that can efficiently radiate at time $t$. A loop of size $\ell$ emits with characteristic wavenumber $k \sim 1/(\alpha\,\ell)$, where $\alpha = \mathcal{O}(1$–$10)$ parametrizes the relation between the loop size and the wavelength of the dominant radiated mode. A representative choice is $\alpha\simeq 4$, motivated by the picture that the relevant mode probes roughly twice the loop diameter in one oscillation. We vary $\alpha$ in the range $1\text{--}10$ to assess the associated uncertainty.

In the relativistic scaling regime, friction is negligible and the largest radiating scale is set by the network morphology, $\ell\sim L(t)$. The morphology-induced cutoff thus corresponds to $x(t) \equiv k/H \simeq 2t/(\alpha \, L(t))$. However, simulations of frictionless networks show an intrinsic suppression of very soft modes with $k/H \lesssim \mathcal{O}(10)$. To account for this suppression, we impose a minimal spectral cutoff $x_{\rm IR}$ and take
\begin{equation}
    x_{\min}^{\rm sc}(t) = \max\left[\frac{2t}{\alpha\,L(t)}, \, x_{\rm IR}\right],
    \label{eq:xmin_scaling}
\end{equation}
which reproduces the absence of power at small $k/H$ in the spectrum of Ref.~\cite{Gorghetto:2018myk}. In the Kibble regime both limitations can be relevant, and we retain the general form
\begin{equation}
    x_{\min}^{K}(t)
    = 2t\,\max\left[\frac{1}{\alpha\,L(t)},\,\frac{1}{\alpha\,\ell_f(t)}\right],
    \label{eq:xmin_kibble}
\end{equation}
so that deep in the friction-dominated phase ($\ell_f \ll L$) the cutoff is friction-limited, while near the transition to scaling ($\ell_f \sim L$) it smoothly approaches the morphology-limited expression in Eq.~\eqref{eq:xmin_scaling}. The ultraviolet bound of the spectrum is set by the core scale $y(t)=m_r/H(t)\simeq 2 f_a t$.

The axion number density at $t_{\rm osc}$ follows from integrating the instantaneous spectrum in Eq.~\eqref{eq:diff_energy_density}. Assuming relativistic emission up to $t_{\rm osc}$, we have
\begin{align}
  n_a(t_{\rm osc})
  &= \int_{t_K}^{t_{\rm osc}} d\tilde{t}\,
     \left[\frac{a(\tilde{t})}{a(t_{\rm osc})}\right]^3
     \int_{\ln k_{\min}(\tilde{t})}^{\ln m_r}
     d\ln k\,
     \frac{1}{k}\,
     \frac{d\dot{\rho}_a}{d\ln k}(\tilde{t})\nonumber\\
  &= \int_{t_K}^{t_{\rm osc}} d\tilde{t}\,
     \left[\frac{a(\tilde{t})}{a(t_{\rm osc})}\right]^3
     \frac{\Gamma(\tilde{t})}{H(\tilde{t})}\,
     \rho_{s,\rm tot}(\tilde{t})\,
    \int_{\ln x_{\min}}^{\ln y} d\ln x \, F(x,y)
  \label{eq:n_final}
\end{align}
where we used that the number spectrum is obtained from the energy spectrum by dividing by the particle energy, $E_k \simeq k$.

\begin{figure}[t!]
    \centering
    \includegraphics[width=0.88\linewidth]{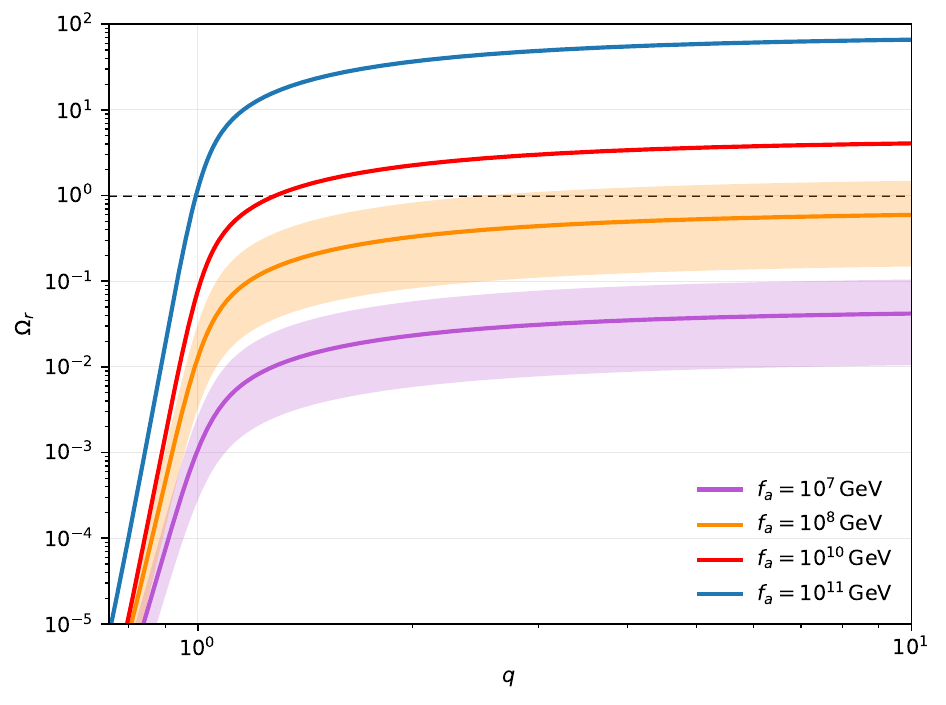}
    \caption{Dependence of the relic abundance ratio $\Omega_r$ (defined as Eq.~\eqref{eq:OmegaR_DM_here}) on the spectral index $q$ for four benchmark decay constants chosen to sample the scaling regime $f_a = 10^{10},\,10^{11}\,\mathrm{GeV}$ and the friction-dominated regime $f_a = 10^{7},\,10^{8}\,\mathrm{GeV}$. The solid lines correspond to the reference choice $\alpha=4$, while the shaded bands show the variation over $\alpha\in[1,10]$. The horizontal dashed line marks $\Omega_r=1$. The curves quickly approach a plateau for $q\gtrsim 1$, indicating that once the spectrum is mildly IR-weighted the abundance is set primarily by the IR cutoff rather than the detailed tilt. For larger $f_a$, the emission is in the scaling regime, where the infrared cutoff saturates to $x_{\rm IR}$ as in Eq.~\eqref{eq:xmin_scaling}, so the result is independent of $\alpha$.}
    \label{fig:q_variation}
\end{figure}

\begin{figure}[!ht]
    \centering
    \includegraphics[width=0.88\linewidth]{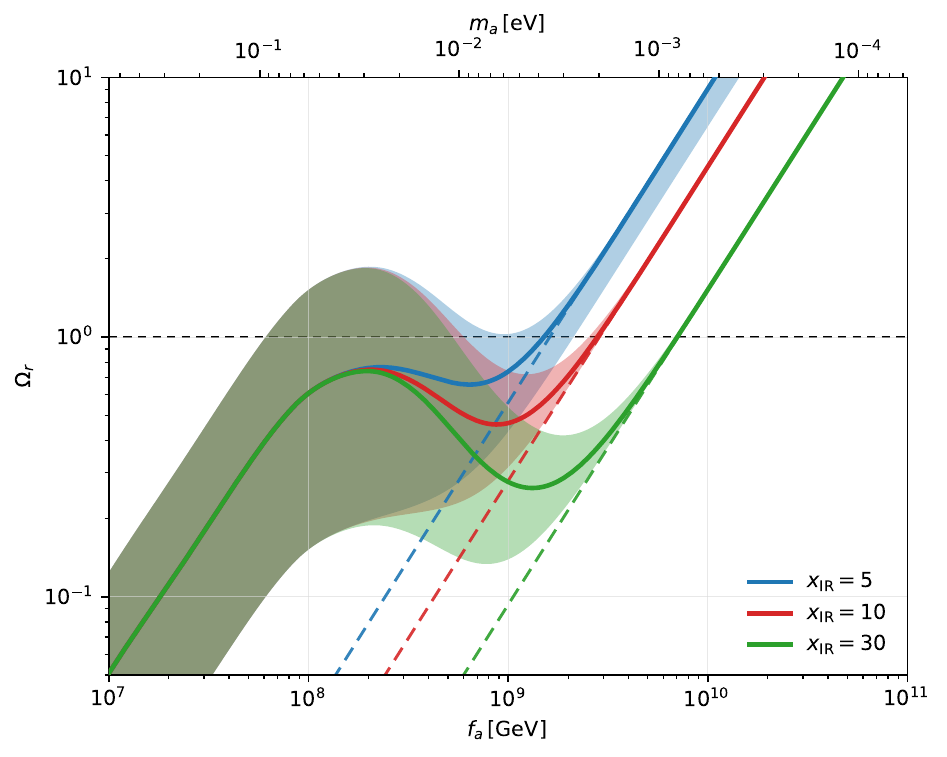}
    \caption{ The relic abundance ratio $\Omega_r$ (defined in Eq.~\eqref{eq:OmegaR_DM_here}) as a function of $f_a$, computed using the spectrum-based estimator for an IR-dominated spectrum ($q \gg $1). The solid curve shows the benchmark value $\alpha=4$, while the shaded band shows the variation over $\alpha \in [1,10]$. The three curves labeled $x_{\rm IR} = 5,10,30$ illustrate that in the friction-limited Kibble regime at low $f_a$, the result is essentially independent of $x_{\rm IR}$, while $x_{\rm IR}$ controls the intercept of the high-$f_a$ scaling branch. The phenomenologically relevant information is the location of the $\Omega_r \simeq 1$ crossings: the high-$f_a$ solution at $f_a \sim \mathrm{few}\times 10^{9}\,\mathrm{GeV}$ in agreement with Ref.~\cite{Gorghetto:2018myk,Gorghetto:2020qws} (with $q>2$) and novel second crossing at $f_a \sim \mathrm{few}\times 10^{8}\,\mathrm{GeV}$, arising from the friction-enhanced axion yield. The dashed lines show the estimate obtained when friction is neglected, while including friction enhances $\Omega_r$ by a factor of order $10$ over much of the friction-dominated regime. For such IR-dominated spectra, the subsequent nonlinearity of the axion potential (discussed in Sec.~\ref{sec:dilution}) will further modify the final abundance (cf. Fig.~\ref{fig:dilution}).}
    \label{fig:GHV_fa_variation}
\end{figure}

The benchmark value $q=0.75$ follows the original frictionless scaling analysis of Ref.~\cite{Gorghetto:2018myk}, which found that a modest UV tilt provides a good description of the spectrum. More recent simulations do not yet agree on a unique spectral shape. Ref.~\cite{Benabou:2024msj} finds an approximately log-flat spectrum with $q\simeq 1$. By contrast, Ref.~\cite{Gorghetto:2020qws} reports a spectral index below unity over the simulated range with a clear growth, well described by a logarithmically varying form $q(\ln x) = q_0 + q_1 \ln x + \cdots$, suggesting that the spectrum becomes increasingly IR-dominated at late times. In view of these differences we present results across a broad range of $q$ to provide an inclusive view of plausible outcomes as shown in Fig.~\ref{fig:q_variation}.

The qualitative behavior of Fig.~\ref{fig:GHV_fa_variation} closely parallels the loop-based case in Sec.~\ref{subsec:method1}. In the extremely IR-dominated limit $q \gg 1$, emission is controlled by the moving infrared cutoff $k_{\min}(t)$, so $\Omega_r$ mainly tracks how this cutoff evolves during axion emission. Three regimes can again be distinguished:\\

\textbf{Scaling regime emission} -- For $f_a \gtrsim 10^9\,\mathrm{GeV}$, the network is already in the relativistic scaling regime by $t_{\rm osc}$, and friction is negligible. The infrared cutoff saturates to the simulation-driven value $x_{\min}(t) \simeq x_{\rm IR}$ independent of $f_a$ during the main emission epoch. The integral in Eq.~\eqref{eq:n_final} is therefore governed by the scaling solution with $\xi(t)\simeq\xi_*$ and $\rho_s\propto f_a^2$. This regime corresponds to the steep, approximately power-law branch of $\Omega_r$ at large $f_a$ in the Fig.~\ref{fig:GHV_fa_variation} \\

\textbf{Correlation length-limited emission} -- For $10^8\,\mathrm{GeV} \lesssim f_a \lesssim 10^9\,\mathrm{GeV}$, the relevant times lie in the crossover between the Kibble and scaling regimes and the IR cutoff is still set by the correlation length $ x_{\min}(t) \simeq 2t/(\alpha \, L(t))$. As $f_a$ decreases from the upper end of this interval, friction becomes important at progressively earlier times, so the network spends an increasingly large fraction of the emission epoch in the Kibble regime. This is visible in Fig.~\ref{fig:GHV_fa_variation} as an increasing departure from the frictionless scaling prediction as $f_a$ decreases. Since the number density at $t_{\rm osc}$ is largely IR-dominated, the enhancement of $\xi(t_{\rm osc})$ over its scaling value in the Kibble regime translates into an increased axion yield as $f_a$ is lowered. As $f_a$ decreases across this intermediate interval, $\Omega_r$ shows a mild increase, instead of the steep power-law fall, characteristic of the pure scaling regime.\\

\textbf{Friction-limited emission} -- For $f_a \lesssim 10^8\,\mathrm{GeV}$, the QCD epoch occurs while the network is deep in the Kibble regime and the strings are strongly friction-damped around $t_{\rm osc}$. The infrared cutoff is friction-limited, $x_{\min}(t) \simeq 2t/(\alpha \, \ell_f(t))$, so decreasing $f_a$ decreases $\ell_f(t_{\rm osc})$, pushing $x_{\min}$ to larger values, reducing the emission of axions. This competes with the increase of $\xi(t_{\rm osc})$ with decreasing $f_a$, yielding a decrease in the effective emitting length per Hubble volume and hence the fall of $\Omega_r$ at the low-$f_a$ end of the plot. However, it is to be noted that the effect of friction is always to increase the number of axions relative to the scaling regime.  \\

The turnover between the three regions again occurs near the point where $\ell_f(t)$ and $L(t)$ are comparable around $t_{\rm osc}$ and where the onset of scaling moves through the emission window.

\subsection{Nonlinear dilution from IR modes}\label{sec:dilution}

At temperatures $T \gg \Lambda_{\rm QCD}$, nonperturbative effects already generate an axion potential, though its impact is initially negligible. In this regime the topological susceptibility $\chi(T)$ is highly suppressed, rendering the axion mass $m_a(T) \propto \sqrt{\chi(T)}/f_a$ negligible and the field evolution effectively free~\cite{Borsanyi:2016ksw}. As the Universe approaches the QCD crossover, the rapid growth of $\chi(T)$ causes the potential to become important at $T_{\rm osc}$. 

The importance of the potential is most apparent for zero-momentum modes, where the bounded potential implies that there is finite energy that can be stored in these soft modes.  Motivated by this, Ref.~\cite{Gorghetto:2020qws} numerically studied what occurs when an excess of energy is stored in the IR modes of the axion. They found that the string network remains dominated by gradient energy as long as the infrared axion energy density satisfies $\rho_{\rm IR}(t) \gtrsim m_a^2(t)\,f_a^2$, where $\rho_{\rm IR}(t)$ denotes the energy stored in the would-be non-relativistic (IR) axion modes. The string network continues to dilute away as radiation until a well-defined later epoch $t_\ell$, defined by $\rho_{\rm IR}(t_\ell) \simeq m_a^2(t_\ell) f_a^2$. Only after this transition does the axion population become non-relativistic, at which point the comoving particle number is conserved and the nonlinear regime concludes.

This mechanism is particularly effective for IR-dominated emission spectra, with spectral index $q>1$, where a large fraction of the power is already in long-wavelength modes and hence in $\rho_{\rm IR}$. For nearly log-flat distributions ($q\simeq 1$) the fraction of energy in these modes is much smaller, so the additional dilution between $t_{\rm osc}$ and $t_\ell$ is correspondingly less pronounced.

To implement this evolution quantitatively, we use the instantaneous emission in Eq.~\eqref{eq:diff_energy_density} and obtain the axion emission spectrum at a time $t$ as
\begin{equation}
    \frac{\partial \rho_a}{\partial k}(t,k) = \int_{t_K}^t dt' \frac{\partial^2\rho}{\partial t' \,\partial k'}(t',k')\,,
\end{equation}
where $k'$ is the physical momentum at the emission time $t'$ and is related to the corresponding physical momentum $k$ at the observation time $t$ by $k = k' a(t')/{a(t)}.$ We define the IR gradient energy at an arbitrary time $t$ as the energy stored in modes whose physical momentum lies between the smallest radiated momentum and a multiple of the axion mass,
\begin{equation}
  \rho_{\rm IR}(t)
  \equiv \int_{k_{\rm min}(t)}^{c_m\, m_a(t)}  dk \,
     \frac{\partial \rho_a}{\partial k}(t,k)\,,
  \label{eq:rhoIR_def_simple}
\end{equation}
where $c_m$ is an $\mathcal{O}(1)$ coefficient fixed by simulations~\cite{Gorghetto:2020qws}.

For the evolution between $t_{\rm osc}$ and the end of the relativistic phase at $t_\ell$, it is convenient to regard the initial spectrum $\partial \rho_a/\partial k(t_{\rm osc},k_{\rm osc})$ and to propagate it forward in time. In this regime the spectrum dilutes away as 
\begin{equation}
  \frac{\partial \rho_a}{\partial k}(t,k)
  = \left[\frac{a(t_{\rm osc})}{a(t)}\right]^{3}
    \frac{\partial \rho_a}{\partial k}
      \Bigl(t_{\rm osc},k_{\rm osc}\Bigr),
  \qquad
  k_{\rm osc} \equiv k\,\frac{a(t)}{a(t_{\rm osc})}\,.
  \label{eq:rho_spectrum_redshift}
\end{equation}

The axion mass grows rapidly around the QCD epoch. For a topological susceptibility $\chi(T)\propto T^{-n}$ with $n\simeq 7$–$8$, one has $m_a(T)\propto T^{-n/2}$. During radiation domination, this implies
\begin{equation}
  m_a(t) \simeq m_a(t_{\rm osc})
    \left(\frac{t}{t_{\rm osc}}\right)^{2}\,,
  \label{eq:ma_growth_t}
\end{equation}
where we take $n=8$~\cite{Gorghetto:2018ocs,Frison:2016vuc}.

The end of the relativistic regime is controlled by the competition between gradient and potential energy in the IR modes. Following Ref.~\cite{Gorghetto:2020qws}, we define $t_\ell$ as the time at which the IR gradient energy first becomes comparable to the potential energy,
\begin{equation}
  \rho_{\rm IR}(t_\ell)
  = c_V\,m_a^2(t_\ell)\,f_a^2\,,
  \label{eq:tl_condition}
\end{equation}
where $c_V=\mathcal{O}(1)$ parametrizes the detailed onset of nonlinearity. Through the dependence of $\rho_{\rm IR}(t)$ on the evolving spectrum and on $m_a(t)$, Eq.~\eqref{eq:tl_condition} defines $t_\ell$ implicitly. In our calculation, $t_\ell$ is obtained by solving Eq.~\eqref{eq:tl_condition} numerically at each point in the $(f_a,\alpha)$ parameter space.

\begin{figure}[t!]
    \centering
    \includegraphics[width=0.88\linewidth]{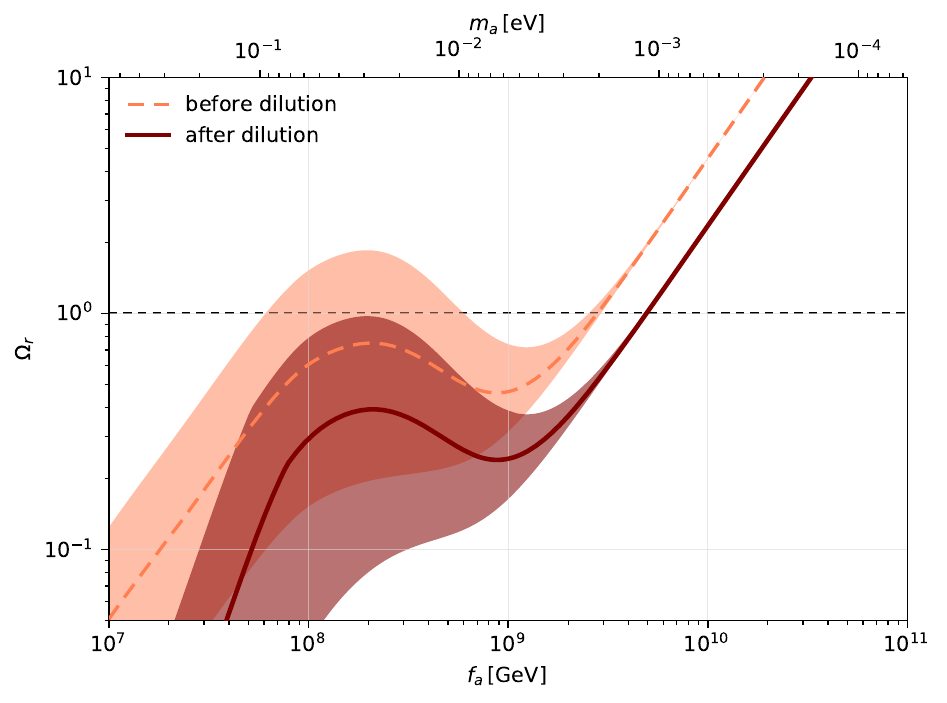}
    \caption{Impact of the post-oscillation dilution on the relic abundance ratio $\Omega_r$ for the IR-dominated spectral estimator with $q\gtrsim2$. The orange band show the result obtained without applying the dilution between $t_{\rm osc}$ and $t_\ell$, while the red band include this effect. The central lines correspond to the reference choice $\alpha = 4$, and the shaded bands span $\alpha \in [1,10]$. Including the dilution suppresses the abundance at smaller $f_a$, reducing the low-$f_a$ branch from $\Omega_r > 1$ to a value of order unity, and shifts the high-$f_a$ $\Omega_r \simeq 1$ crossing to larger $f_a \lesssim 10^{10}\,\mathrm{GeV}$, in agreement with Ref.~\cite{Gorghetto:2020qws}. For the numerical coefficients entering the dilution, we use the simulation-based values of Ref.~\cite{Gorghetto:2020qws}, $c_m = 2.08$, $c_V = 0.13$, and $c_n = 1.35$. }
    \label{fig:dilution}
\end{figure}

Numerical simulations indicate that the nonlinear transient around $t_\ell$ lasts for approximately one Hubble time and the total comoving axion energy is approximately conserved through the nonlinear regime~\cite{Gorghetto:2020qws}. After this transient, the dynamics become linear and the axion field oscillates as a nearly free massive mode. At that point, the comoving axion number density is conserved and can be expressed in terms of the IR energy at $t_\ell$ as
\begin{equation}
  n_a(t_\ell)
  = c_n\,\frac{\rho_{\rm IR}(t_\ell)}{m_a(t_\ell)}
  = c_n\,c_V\,m_a(t_\ell)\,f_a^2\,,
  \label{eq:n_tl_def}
\end{equation}
where $c_n=\mathcal{O}(1)$ encodes the detailed dynamics of the nonlinear regime and is calibrated against the dedicated simulations. For $t>t_\ell$ the axion population is non-relativistic and redshifts as matter. The resulting contribution to the relic density is
\begin{equation}
  \Omega_r \equiv
  \frac{\rho_a}{\rho_{\rm DM}}
  = \frac{m_{a}^0\,n_a(t_\ell)}{\rho_{c,0}\,\Omega_c}\,
    \left[\frac{a(t_\ell)}{a_0}\right]^{3},
  \label{eq:Omega_r_from_tl}
\end{equation}
where $m_{a}^0$, $\rho_{c,0}$, and $\Omega_c$ are defined as in Eq.~\eqref{eq:OmegaR_DM_here}.

In Fig.~\ref{fig:dilution}, we compare the relic abundance obtained from the instantaneous-emission calculation with and without the nonlinear IR dilution. We see a stronger suppression at lower values of $f_a$. The reason for this is that the onset of oscillations occurs earlier when $f_a$ is smaller. Since $t_{\rm osc} \propto f_a^{1/3}$, a smaller $f_a$ corresponds to a smaller $t_{\rm osc}$. Because the oscillations occur earlier, the energy density in axions is larger and stays larger than the axion potential for longer. The end result is that the smaller the $f_a$, the longer the axion redshifts like radiation.

\section{Conclusion} \label{Sec:conclusion}
In this paper, we have shown that friction can drastically modify the dynamics that determine the final axion abundance from an axion string network with  $f_a \lesssim 10^9\,\mathrm{GeV}$. In this window, the network can remain friction-dominated deep into the epoch relevant for determining the final relic abundance, so the usual assumption of an early transition to the relativistic scaling regime need not apply. While the coupling to gauge bosons is typically too small to create significant friction, DFSZ-like models that predict tree-level axion-fermion couplings naturally predict sizable friction from the Aharonov-Bohm effect.

Our semi-analytical framework evolves the long-string network with the full VOS model, which naturally accounts for friction in a certain regime of the dynamics, with the overall normalization fixed to reproduce the scaling value of the long-string density inferred from frictionless simulations, in the appropriate (scaling) limit. We then translate the resulting network history into axion emission using two complementary estimators: (i) a loop-based prescription with a log-uniform instantaneous loop spectrum and friction-limited radiating sizes, and (ii) a spectrum-based prescription formulated in terms of the instantaneous differential power injection with a friction-controlled infrared cutoff. Together, these approaches make explicit how friction delays the onset of scaling, reduces typical velocities near the QCD epoch, and reshapes the infrared part of the emission, thereby modifying the predicted axion yield.

To connect the relativistic emission era to the late-time conserved number density, we also incorporated the nonlinear evolution around the onset of oscillations by tracking the infrared gradient energy until it became comparable to the axion potential energy at $t=t_\ell$, and then mapped the result to the present-day abundance including the associated dilution between $t_{\rm osc}$ and $t_\ell$. We find that friction can significantly enhance the final abundance relative to frictionless treatments, depending on the microscopic drag and infrared modeling, and can open viable parameter space at smaller $f_a$ than is commonly assumed.

The prediction that QCD axion dark matter could arise with $f_a \sim 10^8\,\mathrm{GeV}$, corresponding to $m_a \sim 0.1\,\mathrm{eV}$, is particularly relevant experimentally.  This region of parameter space is much easier to target than the traditional $m_a \sim 100~\mu\mathrm{eV}$ window, and depending on the $\mathcal{O}(1)$ numbers, maybe even be in slight tension with existing constraints~\cite{Ayala:2014pea,XENON:2020rca,Capozzi:2020cbu,Dolan:2022kul}.  It is quite possible that in the near future a QCD axion with $f_a$ this small will be either discovered or robustly constrained~\cite{MADMAX:2019pub,Irastorza:2011gs,ALPHA:2022rxj}.

These results motivate dedicated simulations that include plasma drag and directly compute the emission spectrum and infrared cutoffs in the friction-dominated regime. Aside from refining the prediction for the axion decay constant, such numerical simulations could also verify the VOS prediction of the existence of the Kibble regime. Additionally, a careful treatment of the QCD phase transition and the subsequent production and annihilation of domain walls is warranted. Finally, it would be interesting to further explore various UV completions that contain tree-level axion-fermion couplings.

\section*{Acknowledgments}
The authors would like to thank Abhishek Banerjee for his collaboration during the early stage of the project, as well as insightful comments on the draft. The authors are supported by the National Science Foundation under grant number PHY-2514660 and the Maryland Center for Fundamental Physics.

\appendix
\section{Domain Wall Calculation} 
\label{App:DW}
When the axion mass turns on around the QCD epoch, the discrete shift symmetry selects $N_{\rm DW}$ degenerate vacua and domain walls form. For $N_{\rm DW}=1$, each long string bounds a single wall and the network rapidly annihilates~\cite{Sikivie:1982qv,Vilenkin:1984ib,Benabou:2024msj}. In the thin-wall, sine-Gordon limit
\begin{equation}
  V(a) = m_a^2 f_a^2\big[1-\cos(a/f_a)\big],
\end{equation}
the domain-wall surface tension for $N_{\rm DW}=1$ is the standard result $\sigma = 8\,m_a f_a^2$~\cite{Sikivie:1982qv,Huang:1985tt}. A quasi-circular wall patch of physical radius $r$ thus carries energy
\begin{equation}
E_{\rm wall}(r)\simeq\sigma\,\pi r^2 = 8\pi\, m_a f_a^2\, r^2,
\label{eq:Ewall}
\end{equation}
where we have used the disk area bounded by a circular string loop of circumference $\ell=2\pi r$. The number of axions produced by the collapse of a wall patch is estimated by converting wall energy into particles with typical energy $\langle\omega\rangle\simeq \alpha_w\, m_a$,
\begin{equation}
N_a^{\rm wall}(\ell)
\simeq \frac{E_{\rm wall}(r)}{\langle\omega\rangle}
=\frac{2}{\alpha_w\pi}\,f_a^2\,\ell^2.
\label{eq:Na_per_patch}
\end{equation}
with $\alpha_w=\mathcal{O}(1)$~\cite{Buschmann:2019icd,Buschmann:2021sdq}. 

\subsection*{Sub-horizon wall patches}
If a loop is sub-horizon at $t_{\rm osc}$ (i.e.\ $r<\mathcal{O}(t_{\rm osc})$), the attached wall forms and promptly collapses, converting approximately the entire wall energy into axions within a Hubble time~\cite{Benabou:2024msj}. The number density produced by sub-horizon patches and evaluated at a later time $t_\ell$ is
\begin{equation}
    n_a^{\rm dw,sub}(t_l) = \left[\frac{a(t_{\rm osc})}{a(t_l)}\right]^3
\int^{L(t_{\rm osc})}_0 d\ell \, \frac{dn_\ell(t_{\rm osc})}{d\ell}\,N_a^{\rm wall}(\ell) \, .
\label{eq:ndwsub1}
\end{equation}
Using the log-uniform loop distribution of Eqs.~\eqref{eq:logUniform}, and \eqref{eq:N_of_t}, we obtain
\begin{equation}
    n_a^{\rm dw,sub}(t_l) =\left(\frac{t_{\rm osc}}{t_l}\right)^{3/2}\,\frac{ f_{\rm loop}}{\alpha\pi}\,f_a^2\, \frac{\xi^{1/2}(t_{\rm osc})}{t_{\rm osc}}\,.
\label{eq:ndwsub2}    
\end{equation}
In the Kibble regime, for $f_a \sim 10^8\,\mathrm{GeV}$, this estimate yields $\Omega_r \sim 10^{-5}$, indicating that sub-horizon domain walls make only a subdominant contribution to the axion abundance compared to axions produced by string-loop oscillations.

\subsection*{Superhorizon wall patches}
On superhorizon scales, different parts of the string cannot communicate, so superhorizon segments are effectively frozen in and are conformally stretched. Friction becomes dynamically relevant only once a given mode enters the horizon and the string can respond to drag, at which point its subsequent evolution is governed by the Kibble and scaling regimes.

In principle, the time of horizon entry of these long-wavelength modes provides initial conditions for the friction-dominated dynamics of these superhorizon defects, determining which requires simulations that include plasma drag from the outset. In addition, once these enter the horizon, friction should influence the rate of change of their radius (since it happens through radiation loss). At present, however, large-scale simulations have only been performed in the frictionless case~\cite{Benabou:2024msj,Leite:2011sc,Gonzalez:2022mcx}, so the detailed impact of friction on these initial conditions remains unknown. Nevertheless, frictionless scaling simulations indicate that the contribution of such modes to the final axion abundance is subdominant~\cite{Benabou:2024msj}, and we therefore do not expect the modification of these contributions by friction to significantly alter our relic-density estimates.

\section{Friction on Axion Strings}\label{App:fric} 
In this appendix we give a short self-contained treatment of the drag exerted by a thermal bath on a straight axion string and relate it to the dimensionless drag coefficient $\beta(T)$ that enters the VOS evolution. We closely follow the treatment in Ref.~\cite{Vilenkin:2000jqa}.

\subsection{Kinematic setup}\label{subsec:friction_setup}
Consider an infinite straight string along the $z$-axis, moving with a three-velocity $\mathbf{v}$ in a thermal bath at temperature $T$. In the string rest frame, a bath particle with four-momentum $p^\mu=(E,\mathbf{k})$ has transverse momentum $q \equiv|\mathbf{k}_\perp|$, and scattering in the transverse plane is described by an azimuthal deflection angle $\theta\in[0,2\pi)$. In this frame, the distribution takes the form 
\begin{equation}
    f(\mathbf{k}) = \frac{1}{\exp \big[\gamma\,(E +\mathbf{v} \cdot \mathbf{k})/T\big]\mp 1}\,,
    \label{eq:df_string_frame}
\end{equation}
with the minus (plus) sign for bosons (fermions). For relativistic scatterers we take $E = k$.

The drag force on the string is obtained by summing the momentum transferred from the bath particles to the string per unit time and per unit string length, projected along the direction of $\mathbf{v}$. The flux past the string is controlled by the component of the velocity perpendicular to the string, giving a flux factor
\begin{equation}
    \text{flux per unit length} \propto \frac{q}{E} \simeq \frac{q}{k}\,.
\end{equation}
The microscopic scattering is described by the differential cross section per unit length, $d\sigma/(d\theta\,dz)$. A scattering with deflection angle $\theta$ changes the momentum component along $\mathbf{v}$ by an amount proportional to $k(1-\cos\theta)$. It is therefore natural to introduce the transport cross section per unit length
\begin{equation}
    \sigma_T(q) \equiv \int_0^{2\pi} d\theta\,
    \frac{d\sigma}{d\theta\,dz}\,(1-\cos\theta)\,.
    \label{eq:sigmaT_def}
\end{equation}
Combining these, the force per unit length along the direction of $\mathbf{v}$ can be written as
\begin{align}
    \frac{F}{L} &=
    \int \frac{d^3k}{(2\pi)^3}\, f(\mathbf{k})\,
    \frac{q}{k}\, \left[ k\,\sigma_T(q)\right]\, \nonumber \\
    &=\int \frac{d^3k}{(2\pi)^3}  f(\mathbf{k}) \ \frac{q}{k} \ \int_{0}^{2\pi} d\theta \frac{d\sigma}{d\theta\,dz} \ k \, (1-\cos\theta) \,.
    \label{eq:F_over_L_sigmaT}
\end{align}
Performing the integral, the result can be parametrized as
\begin{equation}
    \frac{F}{L} =
    -\,\beta(T)\,v\,\gamma\,T^3,
    \label{eq:F_def_beta}
\end{equation}
which defines the dimensionless drag coefficient $\beta(T)$. The minus sign indicates that the force opposes the relative motion of the bath and the string. All model dependence is encoded in the differential cross section per unit length $d\sigma/(d\theta\,dz)$ and in the spectrum of species contributing to $f(\mathbf{k})$.

\subsection{Friction due to gauge bosons}\label{sec:App_photon_drag}

In this subsection we specialize the general drag formula of
Sec.~\ref{subsec:friction_setup} to photons, following Ref.~\cite{Agrawal:2020euj}.  Although we focus on the photon, any gauge boson can be accounted for by multiplying by the number of gauge bosons, changing the value of the gauge coupling, and accounting for the extra $1/2$ present in the definition of the axion coupling with $U(1)$ versus $SU(N)$ gauge bosons.  The axion-photon coupling may be written as
\begin{equation}
  \mathcal{L}_{a\gamma\gamma} =
  -\,\frac{e^{2}}{16\pi^{2} f_a}\,a\,F_{\mu\nu}\tilde F^{\mu\nu}\,.
  \label{eq:aFFtilde_photonApp}
\end{equation}
An axion string of unit winding supports a chiral electromagnetically charged zero mode on its core. At energies well below the inverse core size, the resulting anomaly-inflow dynamics can be captured by a bosonized worldsheet
action \cite{Agrawal:2020euj, Witten:1984eb}
\begin{equation}
  S= \int dz\,dt\;
  \bigg[ \frac12\Big(\partial_i\phi - \frac{e}{2\sqrt{\pi}}A_i\Big)^2  - \frac{e}{2\sqrt{\pi}}\phi\,E_z
  \bigg], \qquad i=t,z
  \label{eq:bosonized-action_photonApp}
\end{equation}
where $A_i$ is the pullback of the gauge field to the string and $E_z$ is the electric field along the string.

To compute photon scattering, take the string along $z$ and consider an incident photon whose momentum lies in the transverse plane and whose polarization is along $z$. In the gauge $A_t=A_x=A_y=0$ with $A_z(x,y,t)\equiv A(\mathbf{x}_\perp,t)$, integrating out the worldsheet mode gives the equation of motion as,
\begin{equation}
  \omega^2 A(\mathbf{x}_\perp)=  -\,\nabla_\perp^2 A(\mathbf{x}_\perp) + \frac{e^{2}}{2\pi}\,\delta^{(2)}(\mathbf{x}_\perp)\,A(\mathbf{x}_\perp)\,,
  \label{eq:2D-scattering_photonApp}
\end{equation}
where $\omega$ is the photon angular frequency. The exact solution implies that the field at the core is screened relative to the incident wave
\begin{equation}
  \frac{A(0)}{A_{\rm inc}} \equiv \eta_0(\omega) =
  \frac{1}{1 + \dfrac{e^{2}}{4\pi^{2}}\log\left(\frac{f_a}{\omega}\right)} \, ,
  \label{eq:eta0_photonApp}
\end{equation}
and continuing the scattering solution yields the cross section per unit string length
\begin{equation}
  \frac{d\sigma}{dz} = \frac{e^{4}\,\eta_0^{2}(\omega)}{8\pi^{3}}\, \frac{2\pi}{\omega}\,.
  \label{eq:dsigmadz_photonApp}
\end{equation}
We can now insert this differential cross-section into the general expression Eq.~\eqref{eq:F_over_L_sigmaT} for a thermal photon bath, using the bosonic distribution in Eq.~\eqref{eq:df_string_frame} (two polarizations) and extract the drag coefficient as, 
\begin{equation}
  \beta_\gamma(T) \simeq \frac{4\zeta(3)}{\pi^2} \alpha^2 \eta_0^2(T) \, .
  \label{eq:beta_photonApp_scaling}
\end{equation}
Because it is suppressed by two powers of the coupling, this contribution to friction is negligible.

\subsection{Friction due to fermions}\label{sec:App_fermion_drag}
Fermions that couple derivatively to the axion experience the background of a straight axion string as an effective axial flux tube. In the relativistic limit their scattering off the string is therefore an instance of Aharonov--Bohm (AB) scattering, and we can use the universal AB cross section as our starting point.

In two spatial dimensions, the AB scattering of a charged particle with transverse momentum $k$ off a thin flux tube carrying dimensionless flux $\nu$ has a universal differential cross section per unit length~\cite{Alford:1988sj,Vilenkin:1991zk},
\begin{equation}
  \frac{d\sigma}{d\theta\,dz} =
  \frac{\sin^{2}(\pi\nu)}{2\pi\,k\,\sin^{2}(\theta/2)}\,,
  \label{eq:AB_dsdtheta}
\end{equation}
where $\theta$ is the azimuthal deflection angle in the transverse plane. In AB scattering the incident wavefunction acquires a nontrivial phase $\exp{(i 2\pi\nu)}$ upon encircling the flux tube. The interference of partial waves with this shifted phase produces the characteristic $1/\sin^{2}(\theta/2)$ enhancement at small angles in Eq.~\eqref{eq:AB_dsdtheta}. The overall normalization of the cross section is therefore completely determined by the AB phase and is insensitive to the detailed core structure of the defect.

To apply Eq.~\eqref{eq:AB_dsdtheta} to an axion string, we need to identify the effective flux parameter $\nu$ for a fermion with derivative coupling to the axion. In the background of a straight string along $\hat z$, the axion field
winds once around the core,
\begin{equation}
  a(x) = f_a\,\vartheta, \qquad
  \partial_i a = \frac{f_a}{r}\,\hat{\boldsymbol\vartheta}_i\,,
\end{equation}
where $(r,\vartheta)$ are polar coordinates in the transverse plane. A Dirac fermion $\psi_f$ with axial charge $C_f$ couples derivatively to $a$ via
\begin{equation}
  \mathcal{L}_{af} = \frac{C_f}{2 f_a}\,\partial_\mu a\,\bar\psi_f \gamma^\mu\gamma_5 \psi_f\,.
  \label{Eq: axial coupling}
\end{equation}
To make the analogy with standard AB scattering explicit, it is convenient to rewrite the axion coupling as if it were an axial gauge potential,
\begin{equation}
  A^5_\mu \equiv \frac{C_f}{2 f_a}\,\partial_\mu a\,,
\end{equation}
so that the Dirac operator becomes
\begin{equation}
  i\slashed{\partial}\psi_f
  \,\longrightarrow\,
  \bigl[i\slashed{\partial}-\slashed{A}^5\gamma_5\bigr]\psi_f\,.
\end{equation}
For the winding configuration above, the only non-vanishing component in polar coordinates is
\begin{equation}
  A^5_\theta = \frac{C_f}{2 f_a}\,\frac{\partial_\vartheta a}{r} = \frac{C_f}{2}\,\frac{1}{r}\,,
  \qquad
  A^5_r = A^5_z = 0\,,
\end{equation}
so left- and right-handed fermions couple to equal and opposite azimuthal vector potentials. The axial Wilson phase accumulated by a chiral component on a closed loop around the string is
\begin{equation}
  \Phi_f^{(L/R)} = \pm \oint A^5_\mu dx^\mu =
  \pm \int_{0}^{2\pi} d\theta\,\bigl(A^5_\vartheta\,r\bigr) = \pm \pi C_f\,,
\end{equation}
where the sign is set by chirality. Comparing this to the AB phase
$2\pi\nu$, we see that each chiral component experiences an effective flux
\begin{equation}
  \nu_f^{(L/R)} = \pm \frac{C_f}{2}\,.
\end{equation}
Since the differential cross section depends on $\sin^{2}(\pi\nu)$, both chiralities contribute equally, and the relevant factor is $\sin^{2}(\pi C_f/2)$. Thus the differential cross section per unit length for a given chiral component is
\begin{equation}
  \frac{d\sigma_{f}}{d\theta\,dz} =
  \frac{\sin^{2}\bigl(\frac{\pi C_f}{2}\bigr)}{2\pi\,k\,\sin^{2}(\theta/2)}\,,
\end{equation}
with $k$ the fermion momentum in the transverse plane. 

Using our master formula Eq.~\eqref{eq:F_over_L_sigmaT}, for a Fermi--Dirac distribution one can extract the drag coefficient as 
\begin{equation}
 \beta_{\rm AB}(T)=\frac{3\zeta(3)}{2\pi^2}\, \sum_f g_f(T)\sin^2\left(\frac{\pi C_{f}}{2}\right).
\end{equation}
Here $g_f(T)$ counts effectively relativistic fermionic degrees of freedom at temperature $T$ (including spin, color, and particle-antiparticle). The result is topological in that it depends only on the AB phase (hence on $C_f$), not on the core profile. For order-one values of $\sin^2(\pi C_f/2)$, $\beta_{\rm AB}$ is typically $\mathcal{O}(1\text{--}10)$. In what follows we take $\langle \sin^2(\pi C_f/2) \rangle = 1/2$ as a representative average. For the SM relativistic content above the QCD epoch, summing over the relevant light fermionic degrees of freedom gives $\beta_{\rm AB} \simeq 7.11$.

\subsection{Friction from heavy PQ fermions}\label{app:freeze_out}
In many UV completions such as KSVZ type models, the PQ sector contains heavy fermions $\mathcal{Q}$ with PQ charge and mass $m_{\mathcal{Q}}\gg T$ around the epoch of interest.  If these states were once in thermal equilibrium and then chemically froze out at a temperature $T_F \ll m_{\mathcal{Q}}$, their comoving abundance is fixed thereafter and they form a dilute non relativistic relic. After decoupling, physical momenta redshift as $|\mathbf{k}|\propto T$, so a Maxwellian frozen in at $T_F$ becomes narrower at later temperatures $T$.  In the plasma rest frame we can therefore approximate the phase-space distribution as
\begin{equation}
  f_{\mathcal Q}(\mathbf{k})
  \simeq
  \exp\left[-\frac{m_{\mathcal Q}}{T_F}\right]\,
  \exp\left[-\frac{T_F}{2m_{\mathcal Q}T^2}\,\mathbf{k}^2\right].
\end{equation}
where $\mathbf{k}$ is the physical three-momentum measured in the plasma frame.

To compute the drag on the string, we need this distribution in the string rest frame. Since the relic is non-relativistic and the string speed is small, $v\ll 1$, it is sufficient to work at leading order in the non-relativistic expansion. Boosting by the string velocity $\mathbf{v}$ and writing $\xi$ for the angle between $\mathbf{k}$ and $\mathbf{v}$, we obtain the analogue of Eq.~\eqref{eq:df_string_frame},
\begin{equation}
  f_{\mathcal Q}(\mathbf{k})
  \simeq
  \exp\left[-\frac{m_{\mathcal Q}}{T_F}-\frac{m_\mathcal{Q} v^2}{2T^2}T_F\right]\,
  \exp\left[-\frac{T_F}{2m_{\mathcal Q}T^2}\,\mathbf{k}^2 + \frac{v \, |\mathbf{k}| \cos\xi}{T^2}T_F\right].
\end{equation}

We now specify the long-distance interaction that mediates scattering off the string. At distances large compared to the string core, the coupling of an infinite straight string to the electromagnetic field can be encoded in a local contact term 
\begin{equation}
  \mathcal{L}_{\rm int} \supset
  \lambda\,\delta^{(2)}_{x,y}(\vec{x})\,A_\mu A^\mu\,,
\end{equation}
where $\delta^{(2)}_{x,y}$ is a delta function in the two directions transverse to the string and $\lambda$ is an effective polarizability parameter.  In a concrete model for example, $\lambda$ is generated by integrating out heavy charged degrees of freedom localized on or near the string, but for our purposes it is sufficient to treat $\lambda$ as a low-energy coupling that captures the leading long-distance interaction between the string and photons. Since the relic carries electric charge $e_{\mathcal Q}$, we take the natural estimate $\lambda \sim \alpha_{\mathcal Q}\equiv e_{\mathcal Q}^2/(4\pi)$. The corresponding Feynman rule is
\begin{equation}
  V^{\mu\nu}(q_1,q_2) = 2 i \lambda\,g^{\mu\nu}\,(2\pi)^2
    \delta^{(2)}_{t,z}(q_1 + q_2)\,,
\end{equation}
which enforces conservation of energy and momentum along the string while allowing momentum to be exchanged only in the transverse plane. For the PQ fermion scattering off the string, the amplitude can be written as
\begin{align}
  i\mathcal{M}(q)
  &= \bar u(p_2) \int\frac{d^4 k}{(2\pi)^4}\,
  (-i e_{\mathcal{Q}} \gamma_\mu)\, \frac{i(\slashed{k}+m_{\mathcal{Q}})}{k^2 - m_{\mathcal{Q}}^2 + i\epsilon}\, (-i e_{\mathcal{Q}} \gamma_\nu)\, \frac{-i g^{\mu\alpha}}{(k - p_2)^2 + i\epsilon}\, \frac{-i g^{\nu\beta}}{(p_1 - k)^2 + i\epsilon}\,
  \nonumber \\
  &\hspace{3cm}\times V_{\alpha\beta}(p_1,-p_2) \, u(p_1)
  \label{eq:string_loop_M}
\end{align}
where $q=p_1-p_2$ is the momentum transfer and the $\delta^{(2)}_{t,z}$ factor fixes the longitudinal kinematics, while the loop integral runs over the virtual photon and fermion momenta responsible for transverse momentum transfer.  

The static potential between the string and the non-relativistic fermion at transverse separation $r = |\vec x_\perp|$ follows from the Born approximation as a two-dimensional Fourier transform over the transverse momentum,
\begin{equation}
  V(r)= \int\frac{d^2 q_\perp}{(2\pi)^2}\,
  e^{i \vec q_\perp\cdot\vec x_\perp}\,
  \mathcal{M}(q)\,.
\end{equation}
Using rotational invariance in the transverse plane and expanding at large separations $r\gg m_\mathcal{Q}^{-1}$ gives
\begin{equation}
  V(r) \xrightarrow{r\to\infty}
  \frac{\lambda e_\mathcal{Q}^2}{(2\pi)^3}\,
  \frac{1}{m_\mathcal{Q} r^2}
  \left[ 1 - \frac{2}{3}\,\frac{1}{(m_\mathcal{Q} r)^2} + \mathcal{O}\!\left(\frac{1}{(m_\mathcal{Q} r)^4}\right) \right]
  + \text{oscillatory terms}\,,
\end{equation}
where the oscillatory pieces average to zero on scales large compared to $m_\mathcal{Q}^{-1}$. The leading effect is therefore a repulsive $1/r^2$ potential with overall strength set by $\lambda e_\mathcal{Q}^2/m_\mathcal{Q}$, with $\lambda\sim \alpha_\mathcal{Q}$.  The differential cross-section for a $1/r^2$ potential in 2D is well known to be,
\begin{align}
    \frac{d\sigma}{d\theta\,dz}
    = \frac{\pi \alpha_\mathcal{Q}}{q}\,
    \frac{1}{\big(\theta(2\pi-\theta)\big)^{3/2}}\,,
\label{eq:diff_csx_PQ}
\end{align}
where $q=|\vec q_\perp|$. Using the master formula Eq.~\eqref{eq:F_over_L_sigmaT} and the differential cross-section Eq.~\eqref{eq:diff_csx_PQ}, we get
\begin{equation}
    \beta_{\mathcal{Q}}(T) \sim \frac{\alpha_\mathcal{Q}}{3\pi} \frac{m_{\mathcal Q}^2}{T_F T} e^{-\frac{m_{\mathcal Q}}{T_F}} \, .
\end{equation}
Using the standard freeze-out estimate for heavy KSVZ fermions, the residual abundance carries a Boltzmann suppression
\begin{equation}
  e^{-x_F}\equiv e^{-m_{\mathcal Q}/T_F} \sim \mathcal{O}(10^2)\,\frac{f_a}{M_{\rm Pl}}\,.
\end{equation}
Over the parameter range of interest, the associated friction length $\ell_{f,\mathcal Q}(t)$ (cf.~Eq.~\eqref{eq:lf_def}) always satisfies $2H(t)\gtrsim \ell_{f,\mathcal Q}^{-1}(t)$, so we neglect this contribution in what follows.

\section{The Velocity-One-Scale (VOS) Model}\label{App:VOS}
In this appendix, we summarize the velocity-one-scale (VOS) model for the evolution of a long-string network in an expanding universe, following Refs.~\cite{Martins:1995tg,Martins:2000cs,Copeland:2009ga}. We specialize to the case relevant for axion strings in the post-inflationary PQ scenario, including friction from scattering on the thermal bath, with a logarithmically enhanced tension $\mu \simeq \pi f_a^2 \ln N$.

\subsection{The VOS equations}
The VOS model describes the long string network through the correlation length $L(t)$, defined by $\rho_s(t) = \mu(t) / L^2(t)$, and the RMS string velocity $v(t)$.  In a radiation dominated universe and in the presence of a friction length $\ell_f(t)$, their evolution is governed by
\begin{align}
  \frac{dv}{dt} &=
  (1 - v^2)\left[\frac{k_v}{L} - v\left(2H + \frac{1}{\ell_f}\right) \right], \label{eq:VOS_v_eq_app}
  \\
  2\frac{dL}{dt} &= 2 H L (1+v^2) + c\, v + \frac{L v^2}{\ell_f},
  \label{eq:VOS_L_eq_app}
\end{align}
where $k_v \equiv k(v)$ encodes the average curvature of long strings and $c$ is the loop chopping efficiency that parametrizes energy loss into loops. In some treatments in the literature $k_v$ is treated as a constant phenomenological parameter~\cite{Chang:2021afa,martins2016defect}, which is adequate in the relativistic scaling regime where the network velocity has saturated. However, since we explicitly follow the transition across different dynamical regimes, the velocity dependence of $k_v$ must be taken into account. For this we adopt the phenomenological form motivated by simulations
\begin{equation}
    k(v) = k_0\,\frac{1 - (k_1 v^2)^{k_2}}{1 + (k_1 v^2)^{k_2}}\,,
\end{equation}
with $k_0 \simeq 1.37(7)$, $k_1 \simeq 2.30(4)$, and $k_2 \simeq 1.46(7)$~\cite{Correia:2019bdl,Correia:2020gkj,Chang:2021afa}. There is some debate in the literature regarding the time dependence of the chopping parameter $c$; here we take it to be constant, $c = 0.497$~\cite{Correia:2019bdl,Correia:2020gkj,Chang:2021afa}, and find our results to be insensitive to this choice.

The quantity $\ell_f$ is the friction length arising from drag exerted by the relativistic plasma. For axion strings, it can be written in terms of the drag coefficient $\beta(T)$ defined in Eq.~\eqref{eq:F_def_beta} as
\begin{equation} 
    \ell_f(T) = \frac{\mu}{\beta(T)\,T^3} 
    \simeq \frac{\pi f_a^2 \ln N}{\beta(T)\,T^3}\,.
    \label{eq:lf_def} 
\end{equation}
We define the PQ breaking temperature $T_c$ as the temperature at which the PQ symmetry breaks and denote $t_c$ the corresponding cosmic time by
\begin{equation}
  T_c \equiv f_a, \qquad
  t_c = \frac{M_{\rm Pl}}{2 f_{\rm rad} f_a^2}\,,
  \label{eq:tc_def}
\end{equation}
where $f_{\rm rad} = \pi/3 \sqrt{g_*/10}$. It is convenient to introduce the dimensionless parameter
\begin{equation}
  \theta \equiv \frac{\beta\,M_{\rm Pl}}{2\pi f_{\rm rad}\, f_a}\, ,
  \label{eq:theta_def}
\end{equation}
which packages the dependence on the drag coefficient and PQ scale. This parameter controls the normalization of the Kibble regime and fixes the location of the friction scale relative to the Hubble radius.

For the actual computation of the network solution, we solve Eqs.~\eqref{eq:VOS_v_eq_app}-\eqref{eq:VOS_L_eq_app} numerically rather than relying directly on the asymptotic solutions, with the initial condition $L(t_c) = \xi_{\rm init}^{-1/2}\,t_c$  where $\xi_{\rm init}$ is an $\mathcal{O}(1)$ constant representing the initial long-string density after PQ breaking.\footnote{The results are largely insensitive to $\xi_{\rm init}$, except for extremely small values $\xi_{\rm init} \ll 10^{-10}$ for which the network never enters the Kibble regime.}

\subsection{Asymptotic solutions}
The analytic solutions for each regime commonly used in the literature arise as asymptotic solutions to Eqs.~\eqref{eq:VOS_v_eq_app}-\eqref{eq:VOS_L_eq_app} under controlled approximations and are useful for isolating the underlying physics. Here we summarize each regime, stating the assumptions explicitly and showing how the solutions connect.

\subsubsection*{Stretching regime}
Immediately after PQ breaking, the strings are strongly friction-damped and have very small velocities, so that $v \ll 1$.  The friction length is much shorter than the Hubble radius, and plasma drag dominates over Hubble damping as $\ell_f^{-1}(t) \gg H$.  In this situation the acceleration of the network is slow compared to the curvature-driven acceleration, $\lvert dv/dt\rvert \ll 1/L$, and the velocity is set by a quasi-static balance between curvature and friction. The velocity equation reduces to
\begin{equation}
  0 \simeq \frac{k_v}{L} - \frac{v}{\ell_f}
  \qquad\Rightarrow\qquad
  v_{\rm st}(t) \simeq k_v\,\frac{\ell_f(t)}{L(t)}\,.
\end{equation}
At these early times, loop production and frictional energy loss are still subleading in the length equation. Both $ c\,v$ and $L v^2/\ell_f$ are much smaller than the Hubble stretching term $2 H L$.  Under the same approximations, Eq.~\eqref{eq:VOS_L_eq_app} therefore becomes
\begin{equation}
  2\frac{dL}{dt} \simeq 2 H L
  \qquad\Rightarrow\qquad
  L_{\rm st}(t) \simeq L_c\left(\frac{t}{t_c}\right)^{1/2},
  \label{eq:L_str}
\end{equation}
with $L_c \equiv L(t_c)$ being the initial correlation length at PQ breaking.  In the stretching regime, the network is thus approximately frozen into the expansion, with the long strings simply conformally stretched while the velocity grows slowly from an initially very small value. The stretching regime ends when loop production and friction start to compete with Hubble stretching.

\subsubsection*{Kibble regime}
In these transient times, friction is still important but the strings are no longer purely frozen in.  Curvature and loop production start to play a significant role, while the typical velocity remains small so that we can neglect $\mathcal{O}(v^2)$ terms. Friction remains stronger than Hubble damping, $\ell_f^{-1}(t) \gg H$, and velocity changes slowly, $|dv/dt| \ll v/\ell_f$. The velocity equation takes the same reduced form
\begin{equation}
  0 \simeq \frac{k_v}{L} - \frac{v}{\ell_f}
  \qquad\Rightarrow\qquad
  v(t) \simeq k_v\,\frac{\ell_f(t)}{L(t)}\,.
  \label{eq:v_quasi_static}
\end{equation}
Unlike in the stretching regime, we keep terms linear in $v$ in the length equation, which includes the last term since $L v^2/\ell_f \simeq k_v v$ using Eq.~\eqref{eq:v_quasi_static}. Thus, the length equation reduces to
\begin{equation}
    2\frac{dL}{dt} \simeq 2 H L + (c+k_v)\, v
\end{equation}
Substituting a power-law ansatz yields the standard Kibble-regime scalings
\begin{align}
  v_{\rm K}(t) &\simeq
  \sqrt{\frac{3}{2\theta(k_v+ c)}}\,
  \left(\frac{t}{t_c}\right)^{1/4}\sqrt{\ln N}\,,
  \label{eq:v_Kibble}
  \\[3pt]
  L_{\rm K}(t) &\simeq
  \sqrt{\frac{2(k_v+c)}{3\theta}}\,
  \frac{t^{5/4}}{t_c^{1/4}}\,
  \sqrt{\ln N}\,.
  \label{eq:L_Kibble}
\end{align}
The correlation length grows faster than the horizon while the long string energy density is still controlled by friction, and the velocity increases slowly as $t^{1/4}$. The onset of the Kibble regime, $t_K$, is defined by the matching condition
\begin{equation}
  L_{\rm st}(t_K) = L_{\rm K}(t_K)\,.
\end{equation}
Using Eqs.~\eqref{eq:L_Kibble} and the stretching solution Eq.~\eqref{eq:L_str}, this gives
\begin{equation}
  t_K \simeq
  \left[
    \frac{L_c\,t_c^{-1/4}}{\sqrt{\frac{2(k_v+c)}{3\theta}}\,
    \sqrt{\ln N}}
  \right]^{4/3}
  \propto (\ln N)^{-2/3}\,L_c^{4/3}\,M_{\rm Pl}^{1/3}\,.
\label{eq:tK}  
\end{equation}
For our purposes, the important point is that $t_K$ is fixed once $L_c$, $\theta$ and $\ln N$ are specified, and most of the phenomenology is controlled by the subsequent late-time evolution, so the precise order one coefficient in $t_K$ is not crucial.

The Kibble regime ends once either the velocity is no longer small or friction is no longer the dominant damping mechanism.  In practice, both conditions are met at approximately the same epoch, so it is convenient to define a characteristic time $t_\star$ by the requirement that the microscopic drag rate becomes comparable to the Hubble rate,
\begin{equation}
  \frac{t_\star}{\ell_f(t_\star)} \sim \mathcal{O}(1)
  \quad\Longleftrightarrow\quad
  2H(t_\star) \sim \ell_f^{-1}(t_\star)\,.
\end{equation}
Inserting the explicit time dependence of $\ell_f(t)$ then shows that $t_\star$ scales as
\begin{equation}
  t_\star \propto \frac{\beta^2 M_{\rm Pl}^3} {f_{\rm rad}^3 f_a^4 \ln^2 N}\,,
\end{equation}
up to an order one numerical factor that depends on the precise threshold chosen for the end of friction domination. In our numerical implementation, we use the exact expression obtained from this condition and verify that it correctly reproduces the transition of $\xi(t)$ into the frictionless scaling regime.

\subsubsection*{Scaling regime}
For $t\gg t_\star$ the friction term becomes negligible compared to Hubble friction, and the VOS equations reduce to the frictionless form
\begin{align}
  \frac{dv}{dt} &\simeq (1-v^2)\left(\frac{k_v}{L} - 2 H v\right),\\
  2\frac{dL}{dt} &\simeq 2 H L (1+v^2) +  c\,v\,.
\end{align}
These equations admit a scaling solution with constant velocity, and correlation length proportional to the Hubble radius
\begin{equation}
  L_{\rm sc}(t) = \sqrt{k_v(k_v+c)} \, t, \qquad
  v_{\rm sc}(t) = v_\star = \sqrt{\frac{k_v}{k_v + c}}\,.
\end{equation}
This constant scaling value is precisely what we see in the late-time plateau of the numerical solution $\xi(t)$.

\bibliography{main}{}
\bibliographystyle{unsrt}
\end{document}